\documentclass[12pt]{iopart}
\usepackage{graphicx,latexsym}
\usepackage{bm}
\usepackage{iopams}
\begin{document}
\title{Geometrical effects and signal delay in time-dependent transport at the nanoscale}

\author{Valeriu Moldoveanu}
\address{National Institute of Materials Physics, P.O. Box MG-7,
Bucharest-Magurele, Romania}
\author{Andrei Manolescu}
\address{Reykjavik University, School of Science and Engineering, Kringlan 1, IS-103 Reykjavik, Iceland}
\author{Vidar Gudmundsson}
\address{Science Institute, University of Iceland, Dunhaga 3, IS-107 Reykjavik, Iceland}
\begin{abstract}

The nonstationary and steady-state transport through a mesoscopic
sample connected to particle reservoirs via time-dependent barriers is
investigated within the reduced density operator method. The generalized
Master equation is solved via the Crank-Nicolson algorithm by taking
into account the memory kernel which embodies the non-Markovian effects
that are commonly disregarded.
The lead-sample coupling takes into account the match between
the energy of the incident electrons and the levels of the isolated
sample, as well as their overlap at the contacts. Using a tight-binding
description of the system we investigate the effects induced in the
transient current by the spectral structure of the sample and by the
localization properties of its eigenfunctions.  In strong magnetic
fields the transient currents propagate along edge states.
The behavior of populations and coherences is discussed, as well as their
connection to the tunneling processes that are relevant for transport.

\end{abstract}


\maketitle

\section{Introduction and motivation}

The transport properties of semiconductor structures have mostly been studied
by measuring the steady-state current in response to a constant source-drain voltage drop
applied on some leads that connect to the sample. This current brings relevant information about
resonant tunneling processes. The dependence of the transport coefficients on various parameters
that are varied in the transport experiments (like magnetic field, plunger gate voltages,
tunneling coefficients) revealed effects that are now milestones of nanoscale transport:
Coulomb blockade, Aharonov-Bohm oscillations, mesoscopic Fano effect etc.

On the other hand, there is a growing interest on the electron dynamics inside quantum dot structures
submitted to time-dependent signals applied at the contacts. In this case the system exhibits a more complex
behavior than in the steady-state regime and the quantity of interest is the time-dependent
current in the leads. Recent proposals for coherent control of electron spin in a quantum dot make
use of time-dependent signals\cite{Hanson} and the real-time detection of electron tunneling through a quantum dot electrostatically coupled to a charge detector has been reported.\cite{Gustavsson}
Another example is the pump-and-probe technique proposed by Tarucha {\it et al.}
in order to extract information about the spin relaxation time from transient current measurements.
\cite{Tarucha}

From the theoretical point of view the transient current calculations have been primarily
performed within the non-equilibrium Keldysh formalism. \cite{Jauho,Stefanucci,Mold1,mold2}
Also, an extension of the Lippmann-Schwinger formalism to time-dependent scattering potentials
was presented in Ref.\ \cite{VG}

A characteristic feature of the abovementioned experiments is that the system under study is
in some sense {\it prepared}. More precisely, the chemical potentials of the leads are such
that the first excited state is above the bias window, while the ground state is
embedded in it. Obviously, the current in the leads cannot capture all the details of the dynamics
of electrons in the sample.
A suitable theoretical description of time-dependent transport through mesoscopic systems
on which initial conditions are imposed should therefore focus on the system itself.

The natural formal tool to be used is then the density matrix formalism which was
successfully used in various problems of quantum optics.\cite{Scully}
When adapted to electronic transport
the general strategy goes as follows: i) One starts with several disconnected subsystems, i.e.\ a sample $S$
and some particle reservoirs characterized by different chemical potentials; ii) at instant $t_0$ the
sample is coupled to the reservoirs via a transfer Hamiltonian $H_T$ which can be in general time-dependent;
iii) starting from the quantum Liouville equation for the
statistical operator $W(t)$ that describes the total system one performs a partial trace over the
reservoirs and writes down an integro-differential equation for the reduced density operator (RDO,
defined in Eq.\ (\ref{redrho})).
The latter is called the generalized Master equation (GME) since it contains both diagonal and
off-diagonal elements of RDO. We recall that the usual rate equation describes the evolution of the
diagonal elements of RDO, i.e.\ the populations. In the GME the effect of the
reservoirs on the sample is taken into account through the so called memory kernel which contains an
infinite sum of time-ordered multiple commutators of the type $[H_T,[..[H_T,\rho]]$
and therefore relates the RDO at instant $t$ to its history at previous
times. Otherwise stated, in its general form the equation for the RDO is
non-Markovian. Usually the effect of the leads is taken into account up to the second order in $H_T$
 which at the physical level describes sequential tunneling processes.
The Markov approximation assumes correlation functions in the leads rapidly decaying in time.
As pointed out by Timm\cite{Timm} the characteristic time for the
decay of correlations in the leads is inverse proportional to the applied bias so that in the linear response
regime the Markov approximation could be again inappropriate.

The Born-Markov approximation seems reasonable for steady-state calculations of the current
in the case of a rather large bias but its applicability to
transient regime is not so clear and has been even questioned
recently.\cite{Vaz} In particular for a rapidly varying pulse applied
on the leads or at the contacts one cannot assume that the correlation
functions decay in time. We recall that such a setup is used in
experiments with turnstile pumps.\cite{turnstile1} Also, if one computes higher moments
(e.g.\ noise) non-Markovian effects need to be included.\cite{Thielmann,Braggio}

In view of these considerations the aim of this paper
is on one hand to investigate the time-dependent
transport in mesoscopic structures by solving the GME without using the Markov
approximation. On the other hand we propose an implementation of the generalized master equation which allows
us to take into account the geometry of the sample and uses its spectral properties in order to
set an ``effective'' size for the reduced density matrix to be computed numerically.
We believe that this represents an important step forward because it opens the way to study larger systems
and capture geometrical effects and details about the electron dynamics inside the system itself.

In order to set the general framework we give below a brief survey of several versions of the
RDO method that have been proposed in the context of quantum transport.
In contrast to the quantum optics where the reservoir is a bosonic environment describing the radiation
field, in transport problems the system is coupled to particle reservoirs. Bruder and Schoeller \cite{Bruder}
established a quantum Master equation for the diagonal elements of the statistical operator
by performing a systematic perturbative expansion in powers of $H_T$. Each term in this expansion
corresponds to a tunneling process. Their calculations were primarily focused on the {\it steady state}
regime and emphasized the interplay of sequential tunneling and inelastic cotunneling processes
(the latter ones are described by taking into account fourth order terms in the
transfer Hamiltonian). 
  In the real-time diagrammatic approach \cite{Konig} one writes the Hamiltonian
of the central region in terms of many-body states that have to be computed
in the presence of electron-electron interaction. For few-level systems one can therefore use exact
diagonalization techniques \cite{cotun} in order to investigate elastic or inelastic cotunneling or,
as is done in Ref.\ \cite{Bruder}, to describe the Coulomb interaction within the orthodox model.
This is somehow different from the standard {\it perturbative} calculations within the nonequilibrium
Green-Keldysh formalism \cite{Haug} where the interaction term is included as a two-particle operator in
the Hamiltonian and therefore an interaction self-energy has to be computed.

Later on Gurvitz and Prager \cite{Gurvitz} realized that in the limit of a high bias
the density matrix obeys modified rate equations resembling the Bloch equations.
A common feature of the rate equation approach is that the level broadening due to the coupling to the contacts
should be included 'by hand' in the equations when integrating over energy. In a recent paper \cite{Tokura}
Tokura {\it et al.} investigated interference effects in parallel quantum dot systems in steady-state regime
using both the Keldysh approach and the Bloch equations.
Pedersen and Wacker \cite{Pedersen} developed a scheme that holds for arbitrary bias and goes beyond
the rate equation approach. The matrix elements of the statistical operator are computed within the
Markov approximation in the wide-band limit.
The authors find out that in the steady-state regime the Born-Markov approximation for the RDO and
the non-equilibrium Green's function formalism (NEGF) lead to similar currents, while in the transient regime the
two approaches give different results.
%It would therefore be of some importance to complete this analysis
%by comparing also the non-Markovian version of the reduced density operator method to the NEGF.

Another implementation of the RDO method for the electronic transport in quantum dot systems was presented
by Harbola {\it et al.} \cite{Harbola} Using the Born-Markov approximation and the wide-band limit the authors
have computed both the Fock space populations (FSP) and coherences (FSC) for a two-level noninteracting
quantum dot. The Master equation that is solved in their case describes the projection of the
density operator on the $n$-particle sector of the Fock space; this procedure was introduced by
Rammer {\it et al.} \cite{Rammer} in the context of quantum measurement theory.
It was also shown that within the rotating wave approximation the coherences are decoupled from 
the populations and then the effects of the former on the steady-state currents is not included.
 A thorough comparative
analysis of various GME methods was presented by C. Timm.\cite{Timm} The master equation derived by 
Schoeller {\it et al.} within the real-time diagramatic methods
\cite{Konig} is shown to coincide to the Wangsness-Bloch version of the GME.

In a recent work Vaz {\it et al.} \cite{Vaz} used Laplace transform methods in order to compute
the Redfield tensor that characterizes the memory kernel and presented numerical simulations for
the Fock space coherences in the non-Markovian case. The main statement of this work is the existence
of long-lived FSC even in the steady-state. 
 Finally one should emphasize that the GME method was also employed for studying the effect of a periodically
oscillating signal applied on the sample itself rather than on the leads. \cite{Welack,Li}
In the present work the time-dependence appears only in the transfer Hamiltonian describing the 
contacts between the leads and the sample. 

The paper is organized as follows. Section II describes the model and presents the main equations, some of the
formal details being given in Appendix A. In Section III we present the applications of the method
and discuss the numerical results. Conclusions are summarized in Section IV.

\section{Theory}

\subsection{ The model Hamiltonian}

We consider a mesoscopic sample that is coupled to two leads (particle reservoirs) at the initial
instant $t_0=0$, but decoupled at earlier time. The reservoirs have
different chemical potentials.
We have therefore three subsystems: the two semi-infinite
leads $l=L,R$ (Left and Right), and the central sample $S$.
The transport problem concerns the evolution of this open quantum system for $t>t_0$.

We denote by
$\psi_q^{l}$ and $\varepsilon^l(q)$ the eigenfunctions and eigenvalues of the single particle
Hamiltonians $h_L$ and $h_R$ describing the semi-infinite leads.
The sample Hamiltonian $h_S$ has eigenfunctions $\phi_n$ and eigenvalues $E_n$.
The single particle Hamiltonian of the disconnected system is $h_0=h_L+h_R+h_S$.
We use small letters for these Hamiltonians in order
to distinguish them form their second quantized form which we shall denote below by capital letters.

Our method can be implemented both for continuous or discrete models. In this work we shall
present only the tight-binding case.
The central region is therefore described by a two dimensional lattice Hamiltonian and the leads are modeled as
one dimensional tight-binding chains. The Hamiltonian of the sample in the coordinate representation reads:

\begin{figure}
\begin{center}
\includegraphics[width=0.45\textwidth]{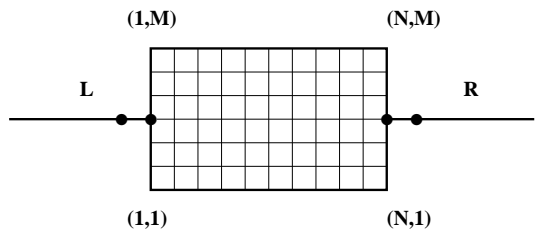}
\end{center}
\caption{A schematic picture of the system we have studied. Two (semi-infinite) one-dimensional
leads are attached to a two-dimensional sample described by a $N\times M$ sites.
The solid dots indicate the contact sites; each site $i$ of the sample is characterized by two indices $(x_i,y_i)$ 
where $x_i=1,..,N$ and $y_i=1,..,M$.}
\label{figure1}
\end{figure}

\begin{equation}\label{Hof}
h_S=\sum_i \lambda_i|i\rangle\langle i|+\sum_{\langle i,j \rangle}
(t_Se^{i\varphi_{ij}}|i\rangle\langle j|+h.c),
\end{equation}
where $i,j$ are nearest neighbor sites in the sample, $\lambda_i$ are on-site energies and
the Peierls phase attached to the hopping parameter $t_S$
describes a constant perpendicular magnetic field.

In order to describe the coupling between the two subsystems we shall add a
perturbation to $h_0$.
We start from the well known single-particle form of the transfer Hamiltonian:
\begin{equation}\label{HTlat}
h_T=\sum_{l=L,R}\chi_l(t)V_l(|0_l\rangle\langle i_l|+h.c),
\end{equation}
where $0_l$ is the site of the lead $l$ which couples to the contact site $i_l$ in the sample
(note that the index $i_l$ is identified once we established a labelling of the sites in the
central region).
 The time-dependent coupling to
the leads is characterized by the switching functions $\chi^l$. Note that in general
$\chi^L\neq\chi^R$ for $t>t_0$ and we are not restricted to the sudden coupling
of the leads as is done in previous works. From the physical point of view time-dependent
couplings can be realized by applied radio-frequency signals to the metallic gates between the
sample and the leads, as is done for example in the turnstile pump experiments. \cite{turnstile}
The constants $V_l$ represent the coupling strength to the $l$-th lead.

Since we are dealing here
with an open system with variable number of particles (recall that the semi-infinite leads
simulate particle reservoirs) it is mandatory to switch to a many-particle Hamiltonian,
although in the present work we shall completely neglect the Coulomb interaction
(we discuss this point further in Section IV).
According to the general rules of
second quantization \cite{Martin} a basis in the Fock space ${\cal F}$ of the coupled system
can be constructed starting from the eigenfunctions $\psi_q^{L/R}$ and $\phi_n$. One
defines creation and destruction operators for electrons in the leads $c_{ql}^{\dagger}$ ($c_{ql}$)
and in the sample $d_n^{\dagger}$ ($d_n$).
Then the second-quantized total Hamiltonian reads as follows:
\begin{eqnarray}\nonumber
H(t)&=&\sum_{l=L,R}\int dq\varepsilon^l(q)c_{ql}^{\dagger}c_{ql}+\sum_n E_nd_n^{\dagger}d_n\\\label{HHat}
&+&\sum_{l=L,R}\sum_n\int dq\chi^l(t) (T^l_{qn}c_{ql}^{\dagger}d_n+h.c),
\end{eqnarray}
where the coefficients
$T_{qn}^{L,R}$ are given by:
\begin{equation}\label{Tqn}
T^{l}_{qn}=V_l{\psi}^{l*}_q(0)\phi_n(i_l),
\end{equation}
The eigenfunctions of the sample are numerically computed while the wave functions $\psi^l_q$
are known analytically (note that they are real and do not depend on the lead index $l$ as we take identical
leads):
\begin{equation}\label{psiq}
\psi^l_q(m)=\frac{\sin(q(m+1))}{\sqrt{2t_L\sin q}},\quad \varepsilon_q=2t_L\cos q .
\end{equation}
In the above equation $t_L$ is the hopping energy of the leads.
Of course one could consider more complicated couplings, taking into account more sites from the central
region coupled to two-dimensional leads. A similar way of constructing coupling matrix elements depending
on junction configuration was proposed in Ref.\ \cite{Maddox}.
The integral over $q$ counts the momenta of the incident electrons such that
$\varepsilon^l(q)$ scans the continuous spectrum of the semi-infinite leads.

The third term in Eq.\ (\ref{HHat}) is the so called transfer or tunneling
Hamiltonian. It has been introduced in the early days of electronic
quantum transport and thoroughly discussed in a series of papers. \cite{Bardeen,Cohen,Caroli,Feuchtwang}
 The tunneling Hamiltonian was extensively used within the
non-equilibrium Green-Keldysh transport formalism.
Usually the wide-band limit approximation is assumed and then the energy dependence of the
coupling coefficients is neglected.
\cite{Haug} More important details of the contacts like position or width
are also omitted.

In the present approach the coefficients $T_{qn}^l$ computed from Eq.\ (\ref{Tqn}) contain
three features: i) The dependence on energies $\varepsilon^l$ and $E_n$ (through $\psi_q^l$ and
$\phi_n$).
ii) The precise location of the contacts between the leads and the sample (i.e. the sites $i_l$.
iii) The probabilities $|\psi_q^l|^2$ and $|\phi_n|^2$ to have electrons at the contact sites.

%We would like to stress that within the reduced density operator approach the geometry of
%the central region could be introduced only in the transfer Hamiltonian.

\subsection{The generalized Master equation (GME)}

Having introduced the second quantized Hamiltonian $H(t)$
we now define the statistical operator of the open quantum system as the solution of the
Liouville equation:
\begin{eqnarray}\label{rho}
i\dot W(t)&=&[H(t),W(t)],\quad W(t<t_0)=\rho_L \rho_R \rho_S, \label{rho_L}
\end{eqnarray}
where
\begin{eqnarray}
\rho_l&=&\frac{e^{-\beta (H_l-\mu_l N_l)}}{{\rm Tr}_l \{e^{-\beta(H_l-\mu_l N_l)}\}}\label{}.
\end{eqnarray}
In the above equation $\rho_S$ is the density operator of the isolated system (that is, for times $t<t_0$)
and serves as an initial condition for the RDO.
$\mu_l$ and $N_l$ denote the chemical potential and the occupation number operator of the lead $l$,
$\rho_l$ being the equilibrium statistical operator of the disconnected lead $l$.
The trace at the denominator is taken in the Fock space of the leads.
The RDO is defined as the (partial) trace on the Fock space of the leads:
\begin{equation}\label{redrho}
\rho(t)={\rm Tr}_L {\rm Tr}_R W(t),\quad \rho(t_0)=\rho_S.
\end{equation}
 The main problem is to find, under suitable approximations, the matrix elements of
$\rho(t)$ with respect to a basis in the Fock space ${\cal F}_S$ of the sample.
 One way to deal with this problem is to compute conditional
reduced operators acting in different $n$-particle sectors of the Fock space (see for example \cite{LiX} ).
Moreover, Li {\it et al.} \cite{LiX} proposed a factorization for the full density matrix
 ($\rho(t)=\sum_n\rho^{(n)}\otimes\rho_{{\rm leads}}$) which generalizes the usual Born-Markov approximation. \cite{Harbola} 
In the present approach we do not impose an equilibrium state on the leads after the coupling is switched on,
which would mean to take $W(t)=\rho_L\rho_R\rho(t)$.  While the steady-state currents are most likely 
not affected by this ansatz the transients are expected to be different when computed within the two approaches.

 We shall use the occupation number basis constructed from the single-particle states $\{\phi_i\}$ of the isolated system.
 Then a many-body state $\nu$ reads as:
\begin{equation}
|{\bf \nu } \rangle = |i_1^{\nu},i_2^{\nu},..,i_n^{\nu}...\rangle,
\end{equation}
where the number $i_n^{\nu}$ indicates if the $n$-th single particle state is occupied ($i_n^{\nu}=1$)
or empty ($i_n^{\nu}=0$). The corresponding energy of the many body state
is denoted by ${\cal E}_{\nu}$ and is given by the sum of the occupied single-particle levels,
i.e ${\cal E}_{\nu}=\sum_n E_ni_n^{\nu}$.
\begin{figure}[tbhp!]
\begin{center}
\includegraphics[width=0.25\textwidth]{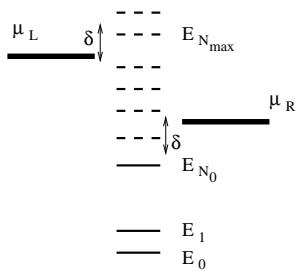}
\caption{The initial configuration in the many-level quantum dot for a given pair of chemical
potentials on the leads $\mu_L,\mu_R$ and a 'gap' $\delta$. The occupied levels are marked by thick lines.}
\end{center}
\label{figure2}
\end{figure}

 It is clear that the size of the reduced density matrix becomes already
very large if the central region accommodates $N\sim20$ electrons and for $N\sim 50$ it seems
quite impossible to compute the entire matrix, even within the Markov approximation. On the other
hand one can easily accept that the number of the many-body states (MBS) that are relevant to
the transport problem is actually much smaller, and at low temperatures is controlled
by the bias applied on the leads. In the present model the bias is included as the
difference between the chemical potentials of the leads i.e.\ $eV=\mu_L-\mu_R$, a procedure which is
also used in the Keldysh formulation of electronic transport. \cite{Jauho,Mold1}

Suppose now that at instant $t_0$ the density operator of the central region is such
that the first $N_0$ single-particle states are occupied and
all the higher states are empty, that is:
\begin{equation}
\rho(t_0)=|\nu_0\rangle\langle {\bf \nu}_0 |,\qquad
|{\bf \nu}_0\rangle = |\underbrace{1,1,....1}_{N_0 \,{\rm states}},0,0,.....     \rangle
\end{equation}
where $\nu_0$ is just the label of the selected many-body state. Moreover, let us consider that the bias
window is fixed such $\mu_R-E_{N_0}\geq\delta$ for a positive $\delta$. When the leads are
plugged to the central region the following scenario is expected: i) The lowest $N_0$ levels
remain occupied and will not contribute to transport as long as the frequency of the coupling signal
$\chi^l$ is small compared to the gap $\delta$; ii) It is reasonable to assume that electrons tunnel
through the dot only via the levels located in the energy range $[\mu_R-\delta,\mu_L+\delta]$; iii)
In the transient regime the occupation numbers of these states will depend on time and will eventually
 settle down in the steady-state regime. Given this setup it is clear that there are only $({N_{\rm max}-N_0})$
 single-particle states which are active in the transport process and consequently it is sufficient to
compute only the matrix elements of the RDO for the ${\cal N}=2^{N_{\rm max}-N_0}$
many-body states having the following form:
\begin{equation}
|{\bf \nu }\rangle =| \underbrace{1,1,....1}_{N_0 \,{\rm states}},i^{\nu}_{N_0+1},....,i^{\nu}_{N_{\rm max}},0,0,.....\rangle
\end{equation}
Let us mention that another interesting initial condition for the density operator
is $\rho(t_0)=|\nu_{{\rm ex}}\rangle\langle\nu_{{\rm ex}} |$ where:
\begin{equation}
|{\bf \nu}_{{\rm ex}}\rangle = |\underbrace{1,1,....1}_{N_0 \,{\rm states}},0,0,..,i_m,0,...     \rangle,
\quad i_m=1
\end{equation}
 namely the one in which besides the lowest $N_0$ occupied levels there is an electron on a higher level
$E_j$ (i.e.\ the initial states is excited). The decay of the state $|{\bf \nu}_{{\rm ex}}\rangle$
as the coupling to the leads evolves could be related to the pump-and-probe experiments in
Ref.\ \cite{Tarucha}.

We proceed now with the equation of motion for the RDO.
It is useful to introduce the notation
$U_0(t,s)=e^{-i(t-s)H_0}$ for the unitary propagator associated to the disconnected system
($H_0=H_S+H_L+H_R$).
 Using the superoperator method developed by Haake \cite{Haake}
we end up with the following GME for the reduced density operator
up to second order in the tunneling Hamiltonian (we give more details of the derivation
in Appendix A):
\begin{eqnarray}\nonumber
\hskip -2cm {\dot\rho}(t)=-\frac{i}{h}[H_S,{\rho}(t)]\\\label{GME}
\hskip -2cm
-\frac{1}{\hbar^2}
{\rm Tr}_L {\rm Tr}_R\lbrace   [H_T(t),\int_{t_0}^t ds [U_0(t,s)H_T(s)U_0(t,s)^{\dagger},
U_0(t,s)\rho(s)U_0(t,s)^{\dagger}\rho_L\rho_R ] ]   \rbrace
\end{eqnarray}
Note that $U_0$ acts on the entire Fock space and cannot therefore be permuted as a whole
inside the partial trace; nevertheless, one can do so for $e^{-i(t-s)(H_L+H_R)}$.
As a next step we rewrite $H_T$ and $H_S$ in terms of many-body states and then
work out the double commutator in Eq.\ (\ref{GME}). Using the completeness relation
$\sum_{\alpha}|\alpha\rangle\langle\alpha|=1$ we have:
\begin{equation}\label{HT2}
H_T(t)=\sum_{l=L,R}\sum_{\alpha,\beta}\int dq\:\chi^l(t)
({\cal T}^l_{\alpha\beta}(q)|{\bf \alpha }\rangle\langle{\bf \beta}|c_q +h.c.),
\end{equation}
where we have introduced a scattering operator ${\cal T}$ acting in the Fock space of the system:
\begin{eqnarray}\label{Toperator}
{\cal T}_l(q)&=&\sum_{\alpha,\beta}{\cal T}_{\alpha\beta}^l(q)|{\bf \alpha}\rangle\langle {\bf \beta}|\\
{\cal T}_{\alpha\beta}^l(q)&=&\sum_nT^l_{nq}\langle {\bf \alpha}|d_n^{\dagger}|{\bf \beta}\rangle.
\end{eqnarray}
It is clear that ${\cal T}_{\alpha\beta}^l(q)$ describes the `absorption' of electrons from the leads to the
system and changes the many-body states of the latter from $\beta\to\alpha$. Note that in the numerical
implementation the index $n$ counts only those single-particle states within the active energy interval
and that in order to have a nonvanishing ${\cal T}$ the number of electrons in the
many-body states $\alpha$ and $\beta$ have to differ by one.

Replacing (\ref{Toperator}) in (\ref{GME}) and using the well known identities:
\begin{eqnarray}\nonumber
e^{itH_l}c_{ql}e^{-itH_l}&=&c_{ql}e^{-i\varepsilon^l(q)t}:={\tilde c}_{ql}(t),\\
e^{itH_l}c_{ql}^{\dagger}e^{-itH_l}&=&c_{ql}^{\dagger}e^{i\varepsilon^l(q)t}
:={\tilde c}^{\dagger}_{ql}(t),
\end{eqnarray}
as well as the correlation functions of the leads ($f_l(\varepsilon^l(q))$ denotes the Fermi function
that characterizes the lead $l$):
\begin{eqnarray}\nonumber
{\rm Tr}_l\{\rho_l{\tilde c}_{ql}(t){\tilde c}^{\dagger}_{kl}(t')\}&=&
e^{-i(t-t')\varepsilon^l(q)}\delta(q-k)(1-f_l(\varepsilon^l(q)))\\\nonumber
{\rm Tr}_l\{\rho_l{\tilde c}^{\dagger}_{ql}(t){\tilde c}_{kl}(t')\}&=&
e^{i(t-t')\varepsilon^l(q)}\delta(q-k)f_l(\varepsilon^l(q))
\end{eqnarray}
one writes the GME into a compact form:
\begin{eqnarray}\nonumber
{\dot\rho}(t)&=&-\frac{i}{\hbar}[H_S,\rho(t)]\\\label{GMEfin}
&-&\frac{1}{\hbar^2}\sum_{l=L,R}\int dq\:\chi^l(t)
([{\cal T}_l,\Omega_{ql}(t)]+h.c)
\end{eqnarray}
where we have introduced two operators:
%\begin{widetext}
\begin{eqnarray}\nonumber
&&\Omega_{ql}(t)=e^{-itH_S} \int_{t_0}^tds\:\chi^l(s)\Pi_{ql}(s)e^{i(s-t)\varepsilon^l(q)}e^{itH_S},\\\nonumber
&&\Pi_{ql}(s)=e^{isH_S}\left ({\cal T}_l^{\dagger}\rho(s)(1-f^l)-\rho(s){\cal T}_l^{\dagger}f^l\right )e^{-isH_S}
\end{eqnarray}
%\end{widetext}
For the simplicity of writing in the above equation we omit to write the energy dependence of the Fermi function
while keeping only the lead index $l$. Equation (\ref{GMEfin}) is the main formal result of the paper.
It leads to a {\it finite} system of ${\cal N}$ coupled integro-differential equations for the matrix elements
$\langle\alpha |\rho(t)|\beta\rangle $ of the RDO. Note that the commutator structure
leads to the conservation of the trace over the sample states, i.e.\ ${\rm Tr}_S {\dot\rho}(t)=0$.
All the tunneling processes of second order in the transfer Hamiltonian are included in Eq.\ (\ref{GMEfin}) and
one can identify loss and gain terms contributing to a given matrix element of the RDO.
Remark that both elastic and inelastic tunneling are taken into account. Electrons in a given state of the
sample are allowed to tunnel out in a state $q$ of the lead at time $t$ and to tunnel back in a different
state at time $t$.  Another important feature is that the sign change of the coupling matrix elements
for subsequent levels is fully taken into account in the above GME. 
 For example one can easily check in the Eq.(\ref{GME})
that terms like ${\cal T}_l\rho{\cal T}_l^{\dagger}f^l$ or ${\cal T}_l^{\dagger}\rho{\cal T}_l(1-f^l)$
contain products of the form $T^l_{mq}(T^l_{nq})^*$ that can have different signs. Such terms
describe processes in which one electron enters the dot on the m-th level while another one
leaves the dot from the n-th level. Their role in transport was recently emphasized by Amir {\it et al.} \cite{Amir}  

At this point one can take further approximations on the GME in order to put it into a
Lindblad form (see for example Ref.\ (\cite{Harbola})). First one applies the Markov approximation
in which $\rho(s)$ is approximated by $e^{i(t-s)H_0}\rho(t)e^{-i(t-s)H_0}$, the time integral is extended to infinity and calculated via a principal value formula. Then one can
take either the rotating wave approximation or the limit of high bias window (i.e.\ $f_{L}=1$ and
$f_R=0$). From the physical
point of view this limiting case means that the electrons can neither flow from the sample to the left lead
nor from the right lead to the sample. In the present work none of these approximations are needed.

Once we have the RDO it is possible to compute the statistical average of the
charge operator in the {\it coupled} sample $Q_S=e\sum_n d_n^{\dagger}d_n$:
\begin{eqnarray}\nonumber
\langle Q_S(t)\rangle &=&{\rm Tr}\{W(t)Q_S\}={\rm Tr}_S\{\rho(t)Q_S \}\\
&=&\sum_n \sum_{\nu} i^{\nu}_n \, \langle\nu | \rho(t) | \nu\rangle \,,
\end{eqnarray}
the traces being now assumed in the Fock space.
Similarly one introduces the charges $Q_{L,R}$ in the leads. We define the net currents in the leads as follows:
$J_{L}(t)=-\frac{dQ_L}{dt}$ and $J_{R}(t)=\frac{dQ_R}{dt}$. We therefore have $J_L>0$
if the electrons flow from the left lead towards the sample and $J_R>0$ if they flow from
the sample towards the right lead. In the transient regime the sign of the net currents
can change. The continuity equation reads
\begin{eqnarray}\nonumber
J(t)&=&J_L(t)-J_R(t)=\frac{d\langle Q_S(t)\rangle}{dt} \\
&=&\sum_n \sum_{\nu} i^{\nu}_n \, \langle\nu | \dot\rho(t) | \nu\rangle \,.
\end{eqnarray}
Using the GME, Eq.\ (\ref{GMEfin}), one can easily identify the contribution of
each level $n$ to the currents in the left and right lead:
\begin{eqnarray} \label{currents}
&& J_l=\sum_n J_{l,n} \nonumber \\
&& J_{l,n}= -\frac{1}{\hbar^2}\sum_{\nu}i^{\nu}_n\int dq\:\chi^l(t)
\langle \nu | [{\cal T}_l,\Omega_{ql}(t)]+h.c. | \nu \rangle  \nonumber \\
\end{eqnarray}
where $i_n^{\nu}=0,1$ specifies if the $n$-th single particle state is occupied or empty.
Observe that the currents are expressed only in terms of the diagonal elements of $\rho$ but this does not
exclude contributions from the off-diagonal elements as well, because all matrix elements are coupled
in the GME.  In order to solve the GME numerically we use the Crank-Nicholson method.
\cite{CN}  The time is discretized and the first derivative of the
density operator with respect to time is evaluated as the mean value of the
forward (or right) and backward (or left) derivatives at the same time point.
The time step was chosen (and tested) to be sufficiently small in order to
capture all physical details of the time evolution of the density operator.
On the RHS of Eq.\ (\ref{GMEfin}) we approximate $\rho(t_{k+1})$ by $\rho(t_k)$ and then
perform iterations until a convergence test is fulfilled. 
The time integration included in the
operator $\Omega_{ql}$ is done recursively: once we know $\rho(t_i)$
for any
$i<k+1$ we update the integral by adding the value of the integrand
corresponding to $\rho(t_{k+1}$.
At any step of the iteration we check the conservation of probability.
It is known that the form of the GME that we use here, in the lowest coupling order, does not guarantee
the positivity of the diagonal elements of $\rho$, {\it i.e.}\ the probabilities of many-body states.
In the numerical simulations presented in Section III we checked the positivity at each time step.
We find that by increasing the coupling constants $V_l$ some of the populations could
take slightly negative values especially in the transient regime. In contrast, for a given sample
the variation of the bias around the active region or a very fast switching of the coupling do not
damage the positivity.

\section{Numerical simulations}

\begin{figure}[tbhp!]
\includegraphics[width=0.45\textwidth]{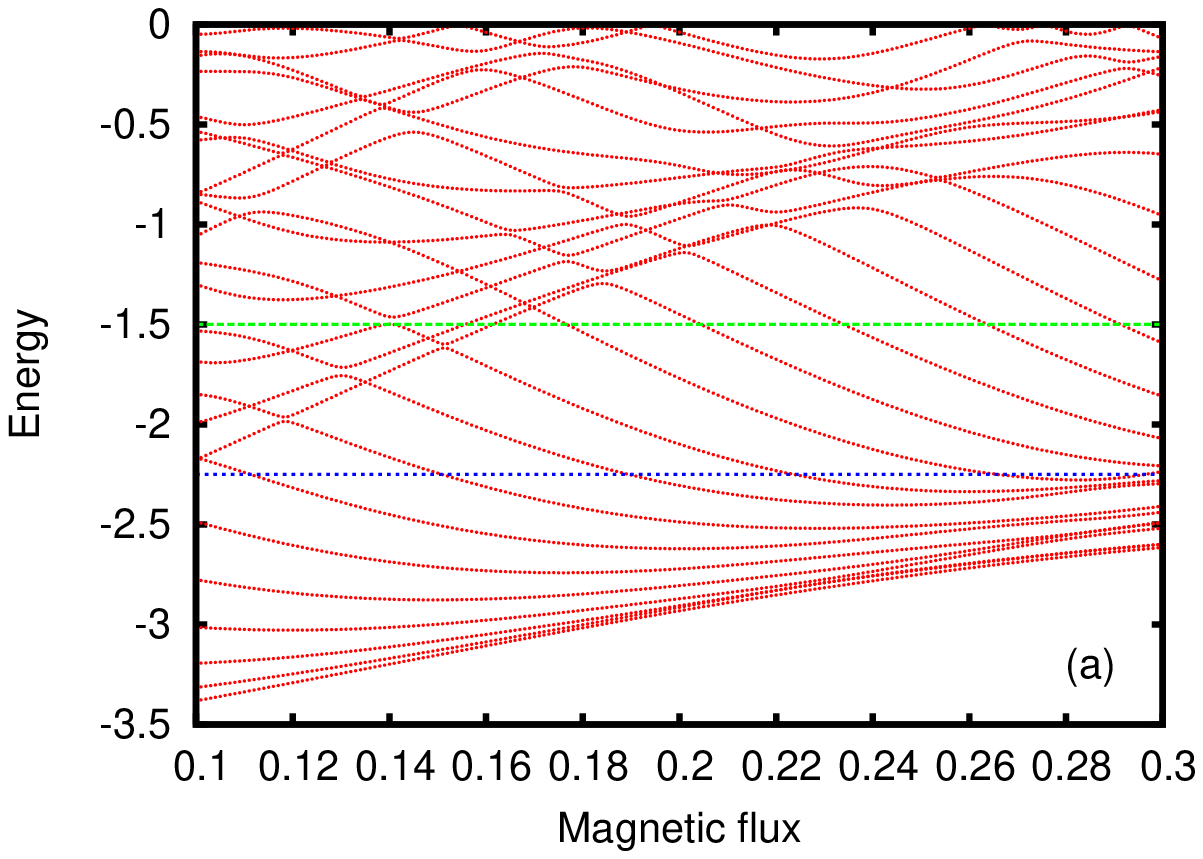}
\includegraphics[width=0.45\textwidth]{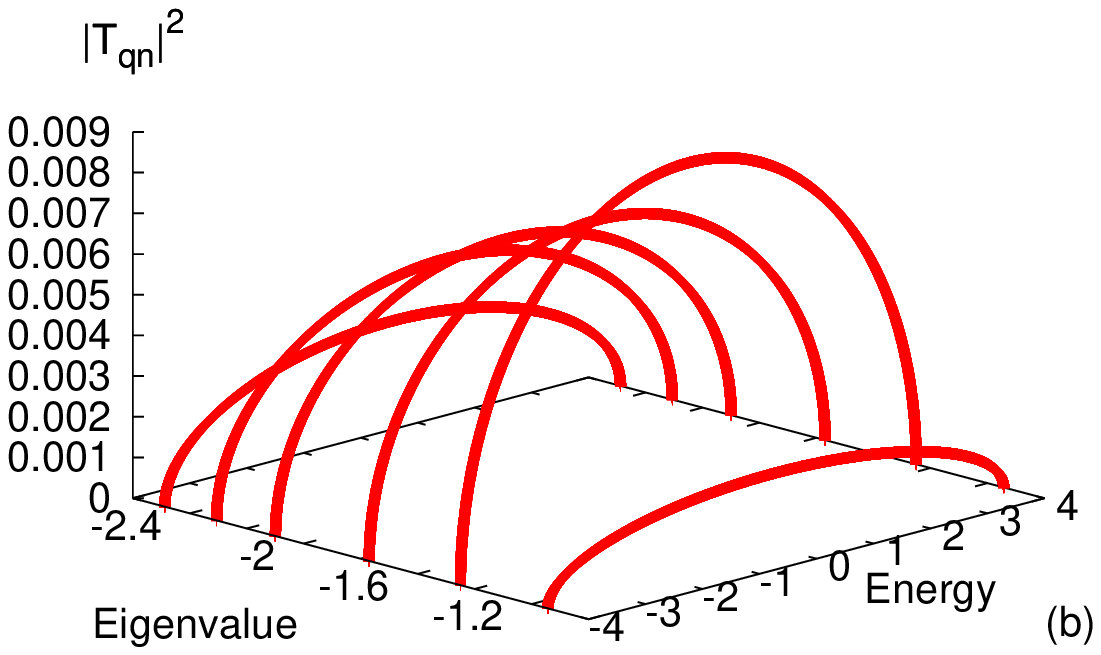}
\caption{(Color online) (a) A part of the Hofstadter spectrum; the two horizontal lines
mark the values of the chemical potentials in the leads, i.e. $\mu_L=-1.5$ and $\mu_R=-2.25$.
(b) The effective coupling between a state $n$ in the relevant window and a state in the
leads having energy $\varepsilon_q$. The maxima correspond to resonant tunneling, that is $E_n=\varepsilon_q$.
Other parameters: $V_L=V_R=1$, $\Phi=0.2$.}
\label{figure3}
\end{figure}

The first sample model is a $5\times 10$ lattice which is large enough to
exhibit the well known Hofstadter spectrum (see Fig\,3(a)) when a strong
perpendicular magnetic field is applied. \cite{Hofstadter}
The Dirichlet boundary conditions
for the two-dimensional discrete Laplacian lead to the formation of edge
states. \cite{H1} The leads are attached at diagonally opposite corners of
the sample.  The magnetic flux is $\Phi=0.2$ and we take $\mu_{L}=-1.5$,
$\mu_{R}=-2.25$.  We consider four active states for transport denoted
by $E_1,..E_4$ and distributed as follows: two states are located within
the bias window, one state above and one state below the window. The
rest of the spectrum is separated from this group of states by a gap
$\Delta E\sim 0.5$. Therefore the transport
properties will be computed from a reduced density matrix which accounts
for 16 many-body states. The temperature is very small, $kT=10^{-4}$.
The switching functions $\chi_{L,R}$ are identical, and describe a smooth
coupling to the leads $\chi_{L,R}(t)=(1-\frac{2}{e^{\gamma t}+1})$. Note that the
parameter $\gamma$ decides how fast we establish the coupling between the two subsystems
($\gamma=1$ if not stated otherwise).
As for the coupling strength we used $V_L=V_R=1$. The important coupling
parameters are actually $T_{qn}^l$ and one can see in Fig.\ 3(b)
that each state is differently coupled to the contact sites.
From Eq.\ (\ref{psiq}). The shape of $|T_{qn}|^2$ as
a function of $\varepsilon_q$ is easily see to be like $\sqrt {4t_L^2-E^2}$

In order to discuss the properties of the RDO we introduce a labeling
of many-body states. It is clear that the only occupation numbers that will be changed when the sample
is coupled to the leads are those associated to the active levels.  The occupation numbers of the
active single-particle states define the many-body state $|{\bf \nu}\rangle$, and
we shall omit the occupation numbers of the frozen (non-active) states.  With four active states we denote
$|{\bf 1}\rangle=|0000 \rangle ,
|{\bf 2}\rangle=|1000 \rangle ,
|{\bf 3}\rangle=|0100 \rangle ,
|{\bf 4}\rangle=|1100 \rangle ,
|{\bf 5}\rangle=|0010 \rangle ,
|{\bf 6}\rangle=|1010 \rangle ,
|{\bf 7}\rangle=|1100 \rangle ,
|{\bf 8}\rangle=|1110 \rangle ,
|{\bf 9}\rangle=|0001 \rangle ,$ etc. (consecutive binary numbers written from right to left).

The interpretation of the sign of the current is: if $J_{L,n}$ and $J_{R,n}$ are
positive the charge flows from the left lead towards the sample and from the sample towards the right lead.
In the steady state we obtain $J_{L,n}-J_{R,n}=0$ for any $n$.
The currents are given in $et_S/\hbar$ units and the time
is expressed in $\hbar/t_S$ where $t_S$ is the hopping energy in the central region; we also take $t_S$
as the energy unit. If one considers an effective lattice constant $a=10$ nm and
the effective mass of GaAs in the definition of the hopping constant $t_S=\hbar^2/2m^*a^2$,
it turns out that the time unit is 1 ps and the current unit is 20 nA.
The hopping energy on leads $t_L=2t_S$, in order to match the spectrum of the 1D lead (i.e $[-2t_L,2t_L]$)
to the spectrum of the central region.

The active region that we consider first contains only the middle 4 states in Fig.\ 3(b).
The remaining two were included in order to to check later on that by taking two more states in the
active region (one below the bias window and one above it) the numerical results are not altered.
The coupling depends on the amplitude of the wave function.  It is clear
for example that a state may not contribute to the current even if it is
energetically inside the bias window if the wave function vanishes
at the contact.

\begin{figure}[tbhp!]
\begin{center}
\includegraphics[width=0.45\textwidth]{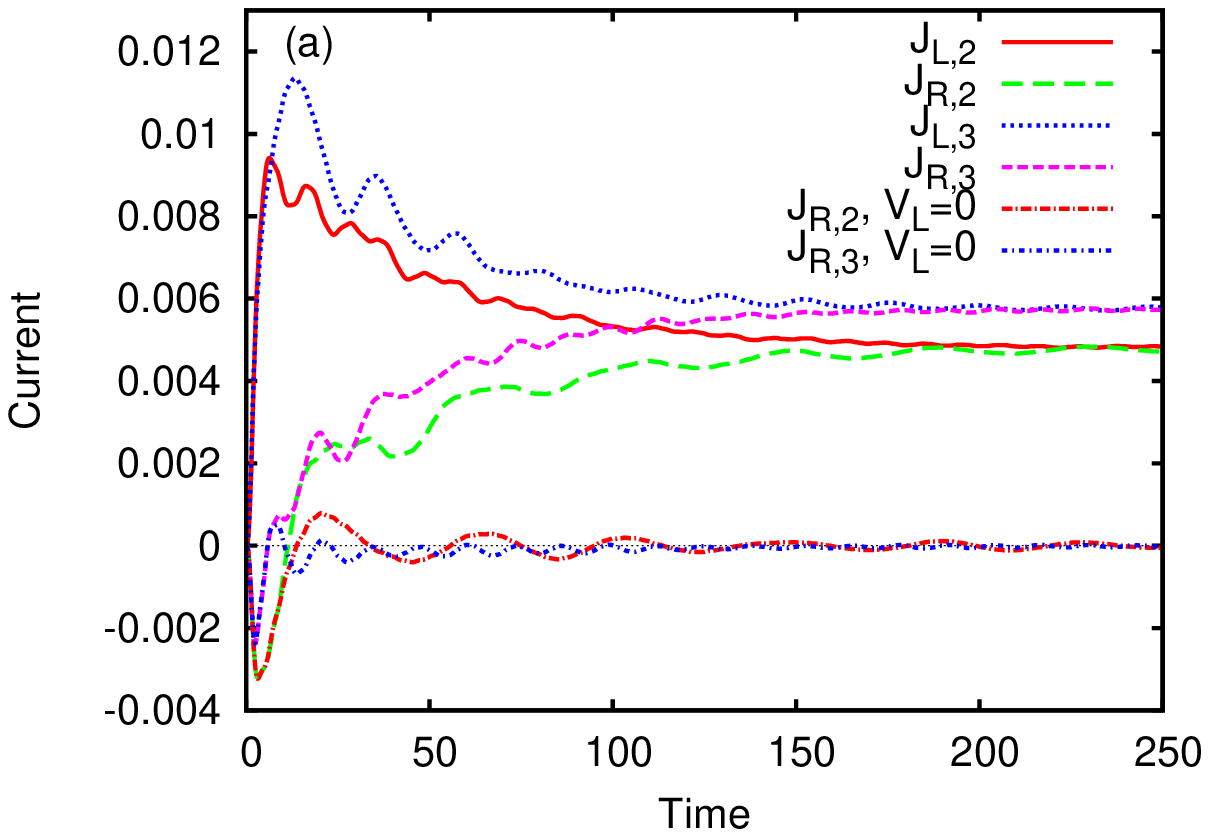}
\includegraphics[width=0.45\textwidth]{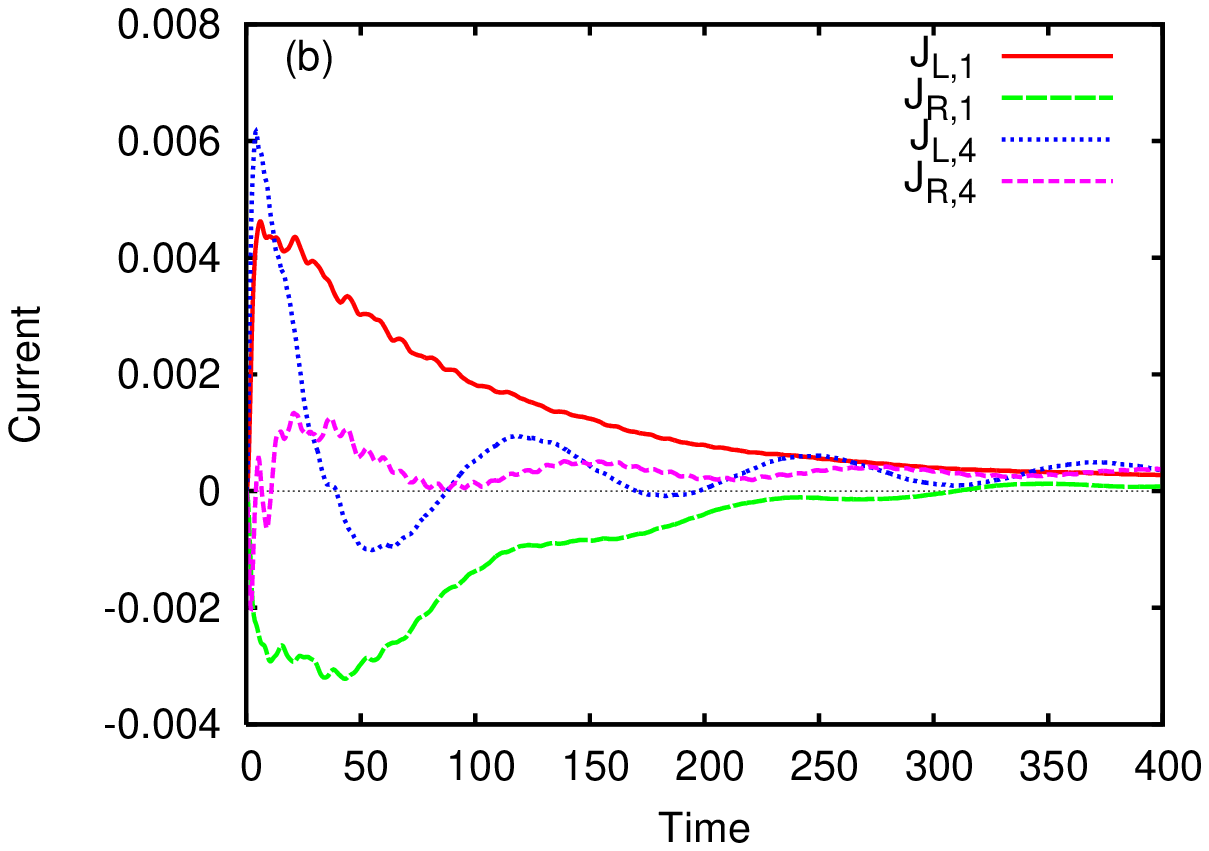}
\includegraphics[width=0.45\textwidth]{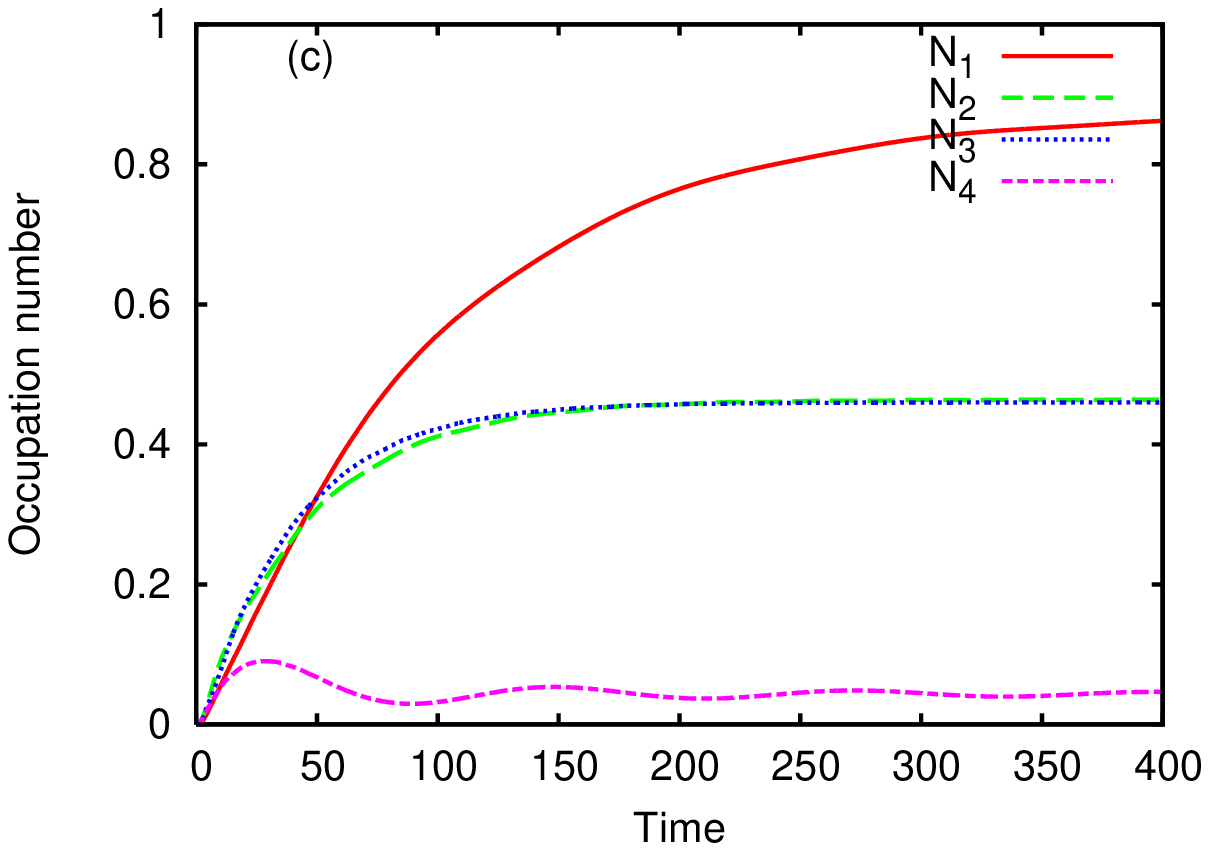}
\end{center}
\caption{(Color online) (a) The currents transmitted through the states below the bias window (2nd and 3rd).
(b) The currents associated to the two states below and above the bias window (1st and 4th).
(c) The average charge on each level within the active region. Other parameters $V=0.75$, $V_L=V_R=1$.}
\label{figure4}
\end{figure}

First we assume that before the coupling to the leads was established the four levels in the active
region were empty, that is $\rho(0)=|{\bf 1}\rangle\langle {\bf 1}|$. The details of the electron dynamics
can be extracted from the currents associated to each level. In Fig.\,4(a) we show and compare the
transients in both leads associated to the two levels within the bias window (see the curves corresponding
to the labels in upper right corner of the figure). In view of further analysis
we include as well the currents in the right lead when the left lead is disconnected  (the two curves
corresponding to the labels in the lower right corner of the figure).
Electrons from both leads can tunnel from or into these states
and the difference between the chemical potentials leads eventually to equal currents in the
steady state. In the transient regime the currents in the two leads behave however
differently: $J_L$ increases abruptly at short times with a bigger slope for
the level whose coupling to the contact is stronger, while the current
flowing into the right lead is delayed. Moreover, this delay depends on the state which carries the current.
 The 3rd level starts to transmit charge earlier (for $t\sim 5$) while the 2nd
needs more time to inject electrons into the right lead  (for $t\sim 20$).
 Fig.\,4(b) gives the currents passing through the 1st and 4th level, which are outside the bias window.
 The following features are noticed: i) The lowest level absorbs charge
from both leads (hence $J_{L,1}>0$ and $J_{R,1}<0$). ii) The currents decrease slowly to zero
giving no contribution to the steady-state current;
iii) The current of the 4th level which is located slightly above the bias window oscillates
with both positive and negative values because in the transient regime this level can gain or
loose charge as well; note however that the amplitude of the oscillation decreases in time.

One can also see that the steady state regime is reached faster by the two states located
within the bias window. This can be better seen in the occupation number of
the corresponding levels which we present in Fig.\,4(c). There are basically two regimes for the
two levels in the bias window.

\begin{figure}[tbhp!]
\begin{center}
\includegraphics[width=0.45\textwidth]{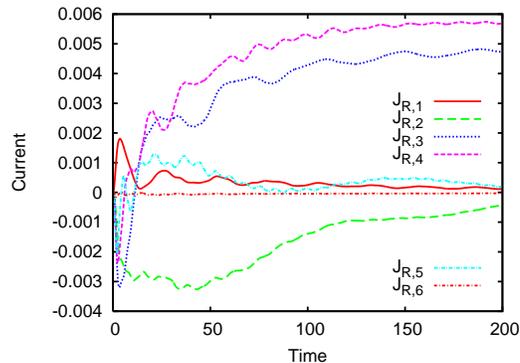}
\caption{(Color online) The current transmitted in the right lead through various states in the
case of an active region containing six levels. Other parameters $V_L=V_R=1$, $kT=10^{-4}$. The current
carried by the 6-th state is vanishingly small and is not given in the figure.}
\end{center}
\label{figure5}
\end{figure}

As the system opens they are charging  at first by absorbing
electrons mainly from the left lead (note that the right lead provides as well a small amount of charge
for a short time); later on the net current in the right lead becomes positive and the steady state
corresponds to a constant occupation number for each level.
This behavior of the occupation numbers is consistent with the numerical calculations we
performed recently using the Green-Keldysh formalism. \cite{Mold1}

The delay of the current in the right lead
which we noticed above could only be associated to the time needed for the electrons
to propagate along the system and tunnel through the contacts. Indeed, when the coupling to the
left lead is switched off the time needed to get a positive (i.e.\ outgoing) current in the right
lead is almost the same as in the two-lead geometry. When $V_L=0$ electrons enter into the sample and spend
some time there before being expelled in the same lead. Nevertheless, in the steady state no current
is generated in the single lead geometry, and the transient oscillations vanish.

As for the lowest level one can see that
in the steady state it is almost full but the time needed to achieve the maximum filling exceed by far the
time needed to have the 2nd and 3rd levels half-filled. This behavior is expected because the lowest level has
the smallest coupling to the leads so that the tunneling processes are slower. We shall see below that plugging the
lead to a contact site where $|T_{q1}|^2$ is larger will accelerate the filling process. Obviously
the {\it total} current will not reach the steady-state unless each partial current does,
even if the corresponding active state is outside the bias window.

Now we take two more single particle states in the active region and compute the transients
by taking the initial state $\rho(0)=|{\bf 2}\rangle\langle {\bf 2}|$ where
$| {\bf 2}\rangle =|100000 \rangle $ (see Fig.\,5). This choice allows a comparison with results given
in Fig.\,4. The 1st and the 6th states carry a vanishingly small current while the remaining
currents are quite similar to the ones in Fig.\,4, which justifies the restriction to 4 single particle states
in the active region. This results could be expected because the couplings $T_{q1}$ and $T_{q6}$
are the smallest one (see Fig\,3(b)).
%We have also compared the numerical results obtained
%with and without the exponential term in the coupling coefficients (not shown).
%It turns out that without the exponential term the transients are slighly larger and may exhibit
%more oscillations, but the general evolution towards the steady-state regime is quite similar
%to the ones obtained with the exponential term.

\begin{figure}[tbhp!]
\begin{center}
\includegraphics[width=0.45\textwidth]{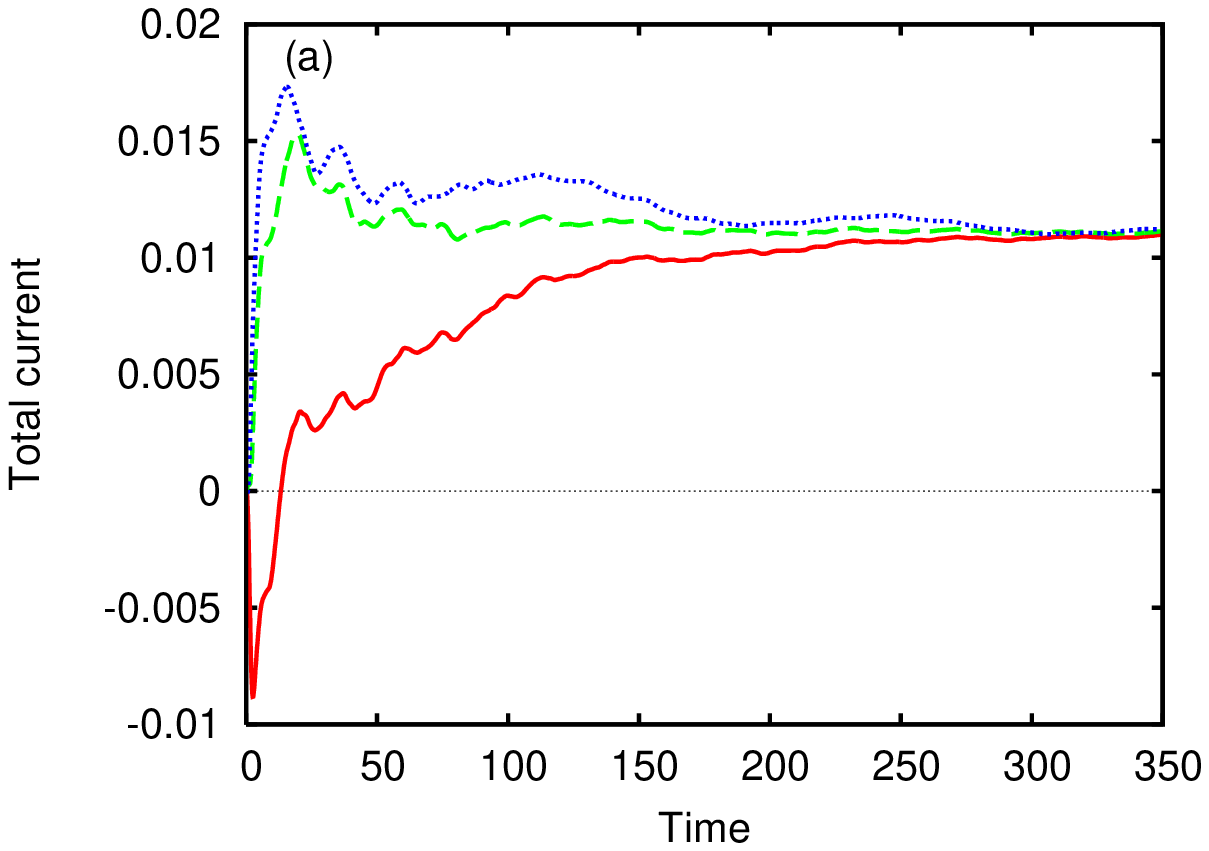}
\includegraphics[width=0.45\textwidth]{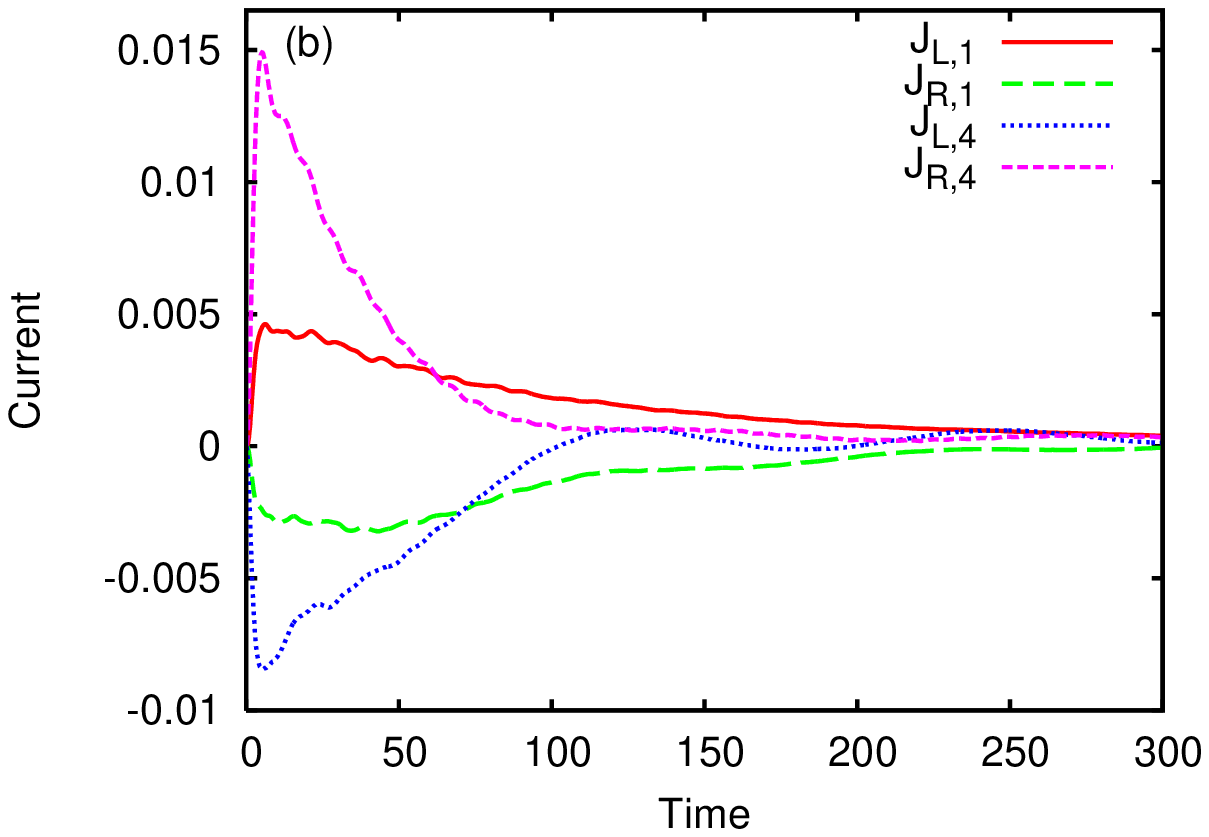}
\end{center}
\caption{(Color online) (a) The total current current in the right lead as a function of time for different initial configurations of
the isolated sample: solid line - $\rho (0)=|{\bf 1}\rangle\langle {\bf 1}|$, long-dashed line -
$\rho (0)=|{\bf 4}\rangle\langle {\bf 4}|$, dotted line $\rho (0)=|{\bf 9}\rangle\langle {\bf 9}|$.
(b) The currents associated to the two states below and above the bias window
when the initial state of isolated system is given by $\rho (0)=|{\bf 9}\rangle\langle {\bf 9}|$.
Other parameters $V_L=V_R=1$, $kT=10^{-4}$.}
\label{figure6}
\end{figure}

The next step of our study is to look at the transients obtained when the levels from the active
interval are already occupied. Here we shall use the main advantage of the reduced density
matrix method, that is,
to take different initial states for the sample. We remind here that in the non-equilibrium
Greens' function formalism the only possible initial state of the disconnected sample is the vacuum; other
initial configurations are not naturally implemented. In Fig.\ 6(a) we compare the total
currents in the right lead obtained for three initial configurations:
$\rho(0)=|{\bf 1}\rangle\langle {\bf 1}|$ (empty system), $\rho(0)=|{\bf 4}\rangle\langle {\bf 4}|$ (the lowest two
levels completely occupied) and $\rho(0)=|{\bf 9}\rangle\langle {\bf 9}|$ (the highest level from the active region
occupied). One notices at once that the steady-state currents do not depend on the initial configuration but the
transients behave differently. First of all, when we start with occupied states in the sample
there is no delay in the onset of a positive current in the right lead, because the electrons immediately tunnel
to the leads; actually, when $\rho(0)=|{\bf 9}\rangle\langle {\bf 9}|$ there is a current flowing to the
left lead as well because the fourth level is located above the bias window. Secondly, the steady state is
achieved faster for the configuration $|{\bf 4}\rangle $ because the lowest level is filled and remains so up to very
small oscillations (not shown). We have checked that the currents $J_{R,3}$ and $J_{R,4}$ coincide with the
ones associated to the initial configuration $|{\bf 1}\rangle $. The relaxation time of the state $|0001\rangle $
can be traced back from Fig.\,6(b) which shows that the currents $J_{R,4}$ and $J_{L,4}$ vanish at $t\sim 100$;
the corresponding occupation number $N_4$ vanishes also there. Since the relaxation process coexists with the
filling process of the level located within the bias window a local minima appears in the total current followed
by a smooth increase towards the steady-state value. The turning point in $J_R$ means that the currents flowing to the right lead through the bias window compensates the decrease in the relaxation current $J_{R,4}$.
More importantly, the decay of this state increases the occupation of the lowest level. This can be seen in Fig.\ 6(b) where the currents entering the lowest level for the configuration $|{\bf 9}\rangle$ is also.
Comparing with the same currents in Fig.\ 4 it is obvious that in this case there is more charge
entering the 1st level and that the currents vanish faster.

We discuss now the behavior of the reduced density matrix. The diagonal elements
reflect the probabilities for having certain occupation numbers in the single particle active states.
It is clear that some  of these configurations are more likely to be realized. Figs.\,7(a) and (b) show
the evolution of some diagonal matrix elements of the RDO.  Again, we start with the
sample in the vacuum state, $\rho(t_0)=\rho_{11}$, where
$\rho_{11}$ is the probability to have all the levels from the active region empty.
It decreases from the initial value to zero because the levels are populated as the coupling to the leads
strengthen; the numerical data suggests an exponential decay.

One can identify several regimes for the populations. At short times
the most favorable nontrivial populations contain just one electron.
Otherwise stated, the system spans the single-particle sector of the Fock space.
Physically this means that two levels cannot be simultaneously occupied shortly
after the coupling is switched on. The occupation probability
of levels with the lowest energies is expected to be higher,
i.e.\ $\rho_{33}>\rho_{55}$. The lowest levels absorb charge from both leads,
while the higher ones are populated mostly due to tunneling
from the left lead. This is true in our case only after a short time, but
because of the energy dependent coupling we are using, after a longer time we see the opposite result.
In a second regime the system is more likely to be in a state from the two-particle sector of the Fock space, namely
$|{\bf 4}\rangle=|1100\rangle $ and $|{\bf 6}\rangle=|1010\rangle $ (Fig.\,7(b)).
The corresponding diagonal elements
$\rho_{44}$ and $\rho_{66}$ increase with time. At even later times the system can accommodate three electrons on the
first levels so $\rho_{88}$ increases as well. In the steady-state regime four configurations have
a significant probability: $|1100\rangle$, $|1010\rangle$, $|1000\rangle$ and $|1110\rangle$. It is obvious
that by pairing these states one recovers the tunneling processes that contribute to the steady-state
currents in the leads. For example, switching from $|1100\rangle\to |1000\rangle$ implies that one electron tunnels
out from the 2nd level in the active region.

\begin{figure}[tbhp!]
\begin{center}
\includegraphics[width=0.45\textwidth]{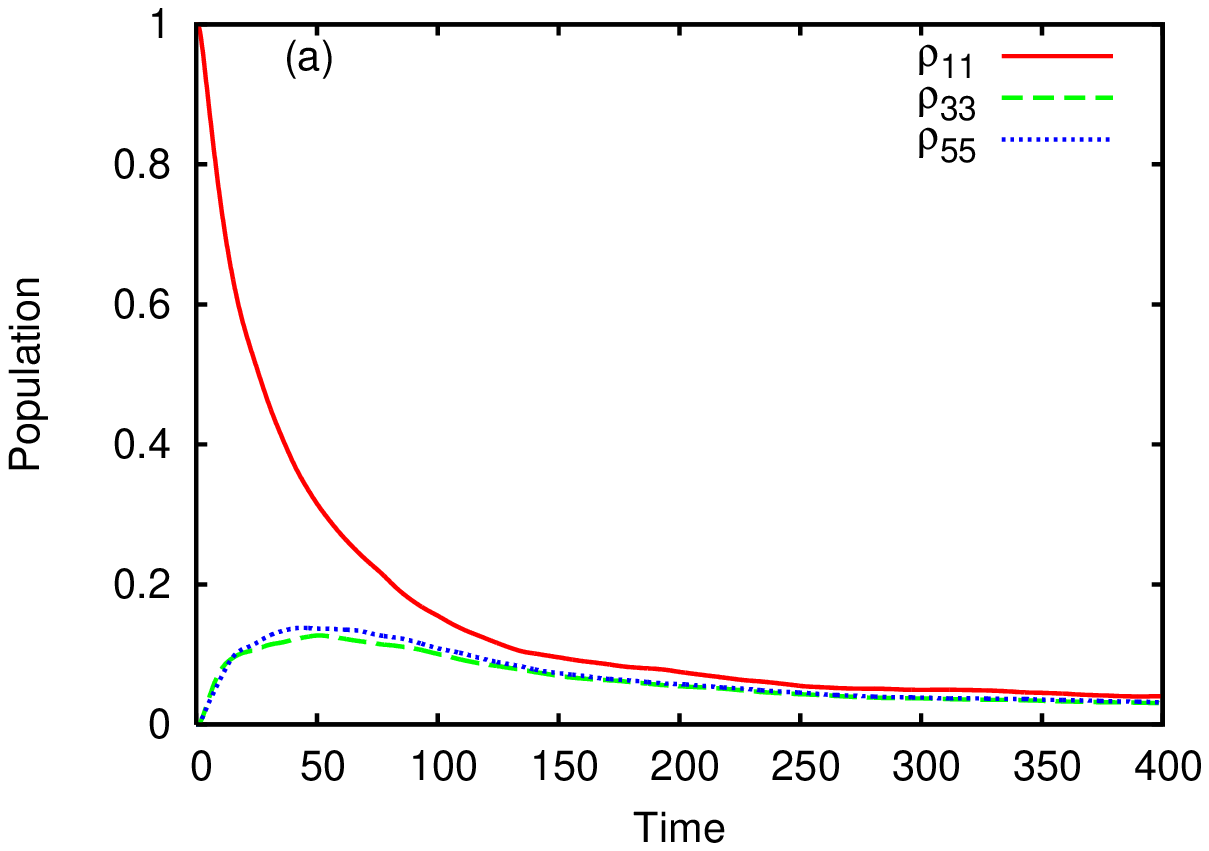}
\includegraphics[width=0.45\textwidth]{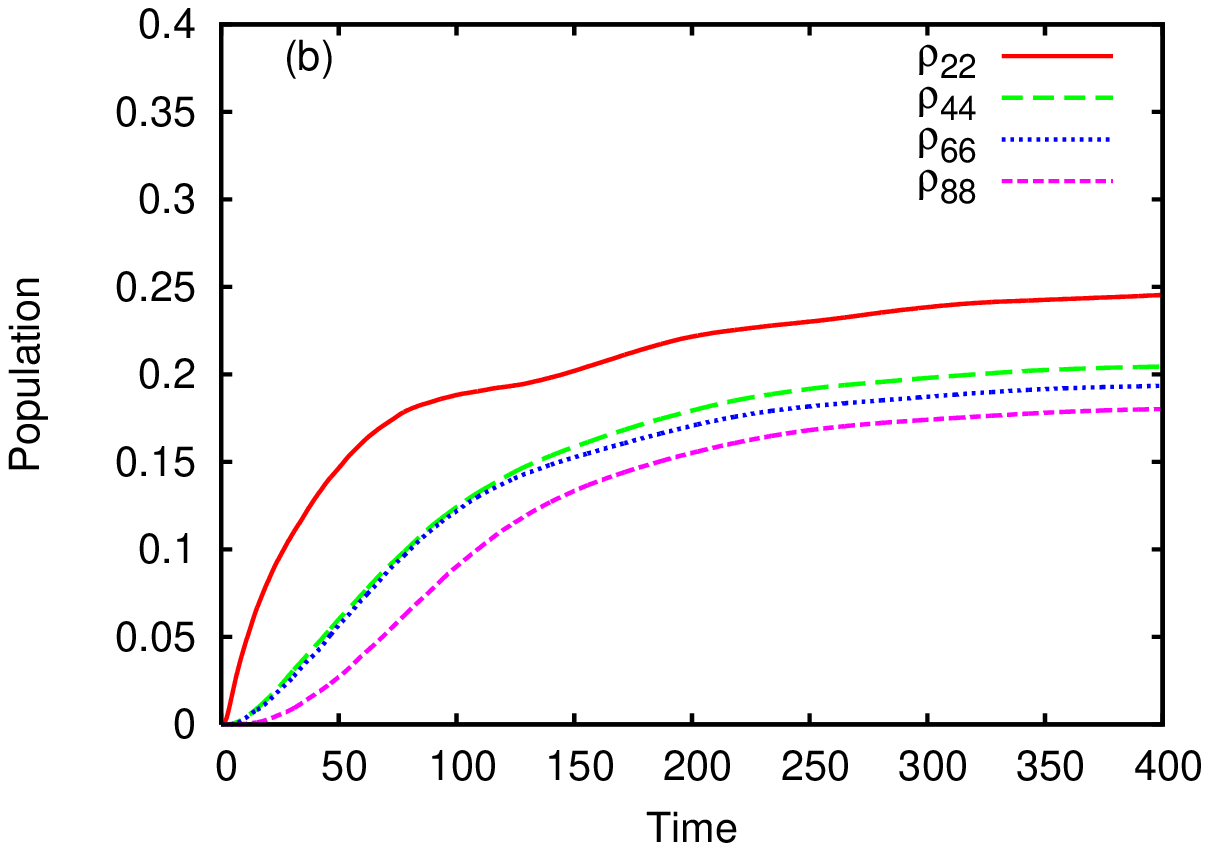}
\end{center}
\caption{(Color online) Time-evolution of the most relevant populations for the 5x10 sites system.
(a) Single-particle configurations decay in the long time limit. (b)
The populations that do not vanish in the steady state (see the text). The other parameters are as
in Fig.\,6.}
\label{figure7}
\end{figure}

The non-Markovian nature of the system implies that off-diagonal elements (the so called coherences)
of the RDO could develop in time, even if we start from a diagonal
density operator at $t=0$. A nonvanishing coherence $\rho_{\alpha\beta}(t)$ means that at instant
$t$ the state of the system cannot be completely described by occupation probabilities.
The coherences can be shown to vanish in the steady-state if, on top of the Born-Markov approximation, one also takes the rotating wave approximation.\cite{Harbola} This is not the case in the GME
case and moreover,
the diagonal and off-diagonal elements of the density operator are coupled.
Note however that the second order approximation with respect to the transfer Hamiltonian
that we keep here implies that coherences between many-body
states with different particle numbers are excluded. This is because
${\rm Tr}\{ \rho_L c_{ql}^{\dagger}c_{ql}^{\dagger}\}={\rm Tr}\{ \rho_L c_{ql} c_{ql}\}=0$.
We show in Figs.\ 8(a) and (b) the behavior of the nonvanishing coherences which are complex quantities
and satisfy the relation ${\rho}^*_{\alpha\beta}=\rho_{\beta\alpha}$.
In general both the imaginary and real parts have oscillations. Nevertheless,
some matrix elements $\rho_{\beta\alpha}$ vanish in the steady state (see Fig.\ 8(b))
while some settle down to a non-vanishing value and contribute indirectly
to the current, via the diagonal elements $\rho_{\alpha\alpha}$. We notice also that coherences do not appear
simultaneously and their oscillations behave differently. For instance, $\rho_{23}$ starts
earlier than $\rho_{46}$ and its oscillations decrease in time; in contrast, although $\rho_{46}$ exhibits
mild oscillations at short times it gradually increases and clearly exceeds $\rho_{23}$ in the steady state.
The explanation of this behavior lies in the dynamics of the occupation numbers in the sample.
On one hand, the state $|{\bf 3}\rangle=|0100\rangle $ is
realized with probability $15\%$ at short times only (see Fig.\ 7(a)) but then this state become less probable.
This is why $\rho_{23}$ decreases at $t\sim 125$. On the other hand both states that appear in $\rho_{35}$
have higher probability ($\sim 20\%$) in the long time limit. Similar arguments explain why the
other two coherences shown in Fig.\ 8(b) vanish in the steady state: they imply sequences of occupation
numbers that are not expected in the long time limit. Moreover, since the probability to find the
system in the state $|{\bf 9}\rangle$  decreases much faster that the one associated to the state
$|{\bf 3}\rangle$ it is clear that $\rho_{95}$ decays well before $\rho_{35}$.

\begin{figure}[tbhp!]
\begin{center}
\includegraphics[width=0.45\textwidth]{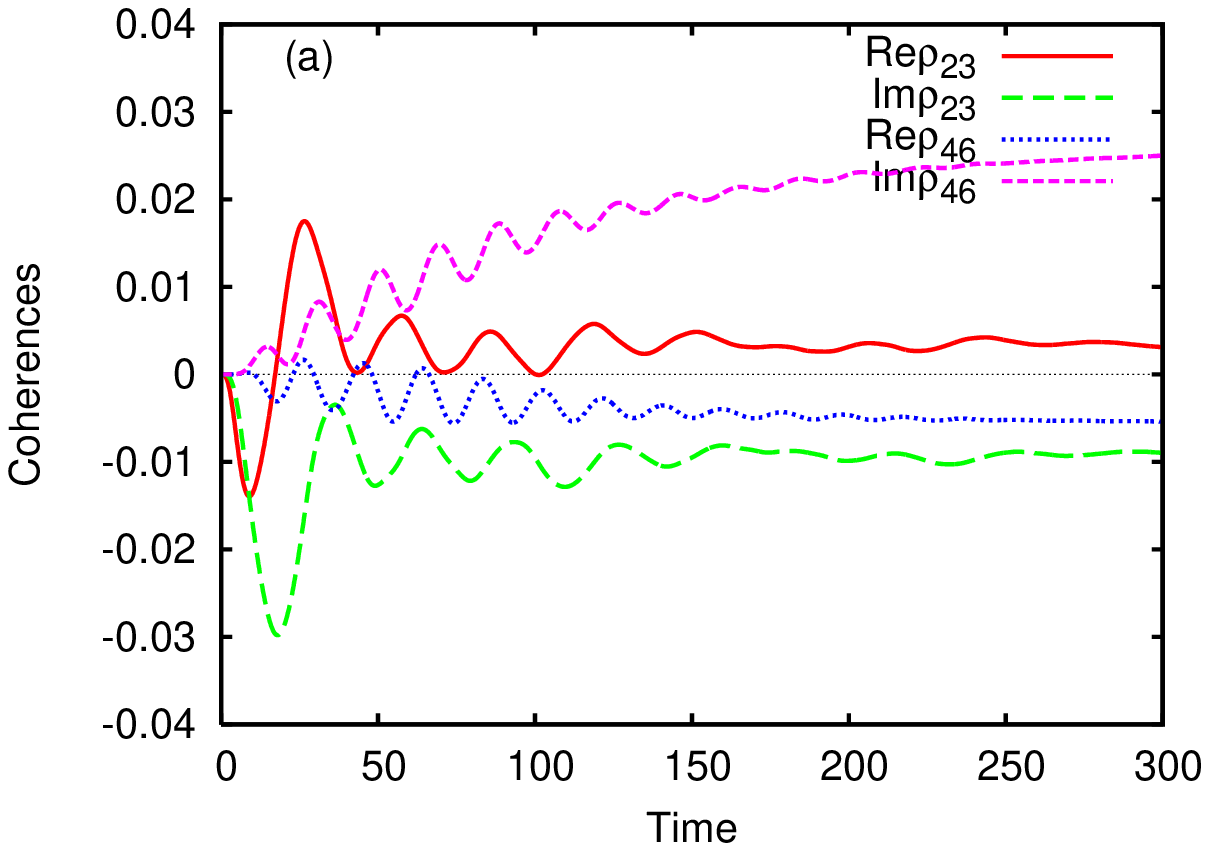}
\includegraphics[width=0.45\textwidth]{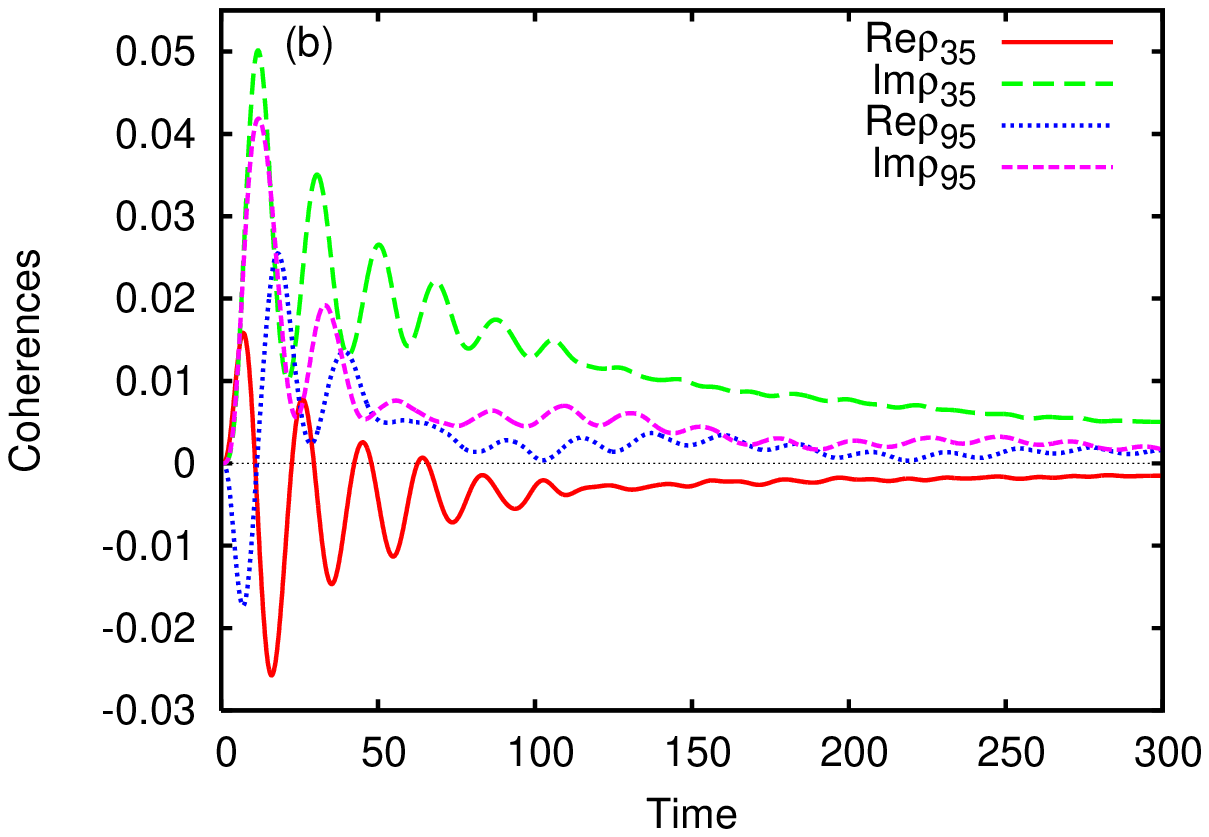}
\end{center}
\caption{(Color online) Time-evolution of the most relevant coherences.
(a) Non-vanishing off-diagonal elements reaching a steady-state value. Remark that $\rho_{23}$ develops
earlier than $\rho_{46}$. (b) Oscillating coherences that vanish in the steady state and therefore do not
contribute to the total current.}
\label{figure8}
\end{figure}

The definition of the coupling coefficients $T_{qn}^{L,R}$ (see
Eq.\,(\ref{Tqn})) implies that by choosing different contact regions
one could obtain different currents as for any $q,n$ the associated
eigenfunctions may overlap differently.  In Fig.\ 9(a) we show
$|T_{q1}|^2$ and $|T_{q3}|^2$ for two setups: the one we already
discussed, where the leads are placed at diagonally opposite corners
of the sample and a second configuration in which the opposite middle
sites are used as contacts, like in Fig.\ 1. The differences are obvious:
in the 2nd configuration the coupling to the
 3rd level decreases and the coupling to the 1st level increases. The
 differences induced in the associated
currents are identified in Fig.\ 9(b): The contribution of the 3rd level
to the current transmitted in the right lead reduces considerably in the
2nd configuration; also the steady-state value of the occupation number
$N_3$ is around 0.8 (not shown) which suggests that this level charges
more than in the 1st setup ($N_3\sim 0.5$ in the steady-state). On the
other hand the current $J_{L,1}$ increases in the 2nd configuration and
the occupation number $N_1$ goes to the steady-state value faster that
in the previous geometry, meaning that the filling of the 1st level is
done easier since the opening to the contacts is higher.

\begin{figure}[tbhp!]
\begin{center}
\includegraphics[width=0.45\textwidth]{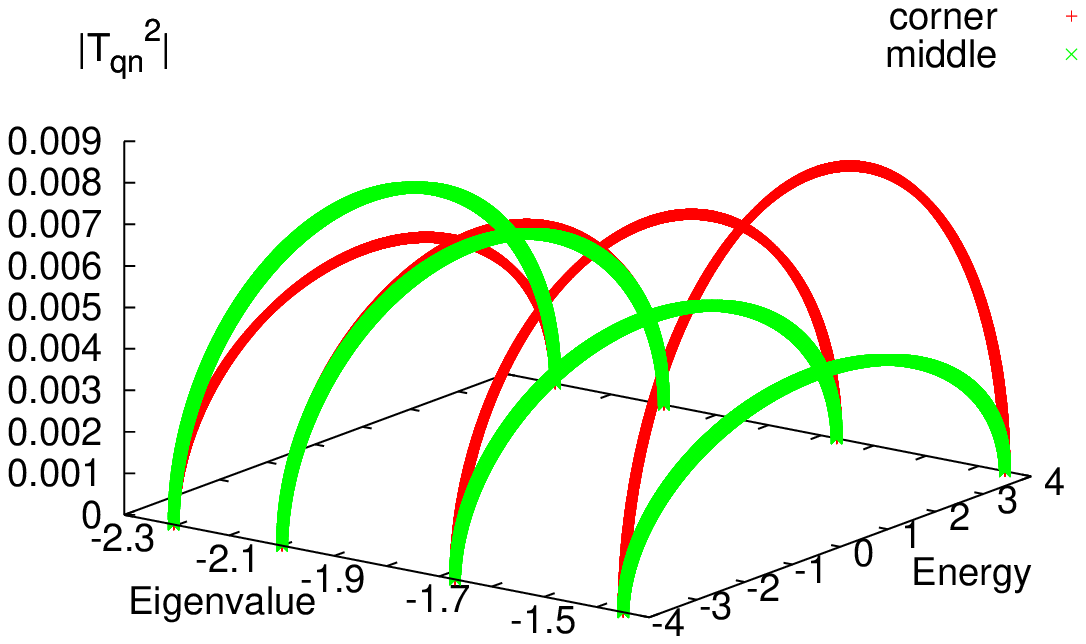}
\includegraphics[width=0.45\textwidth]{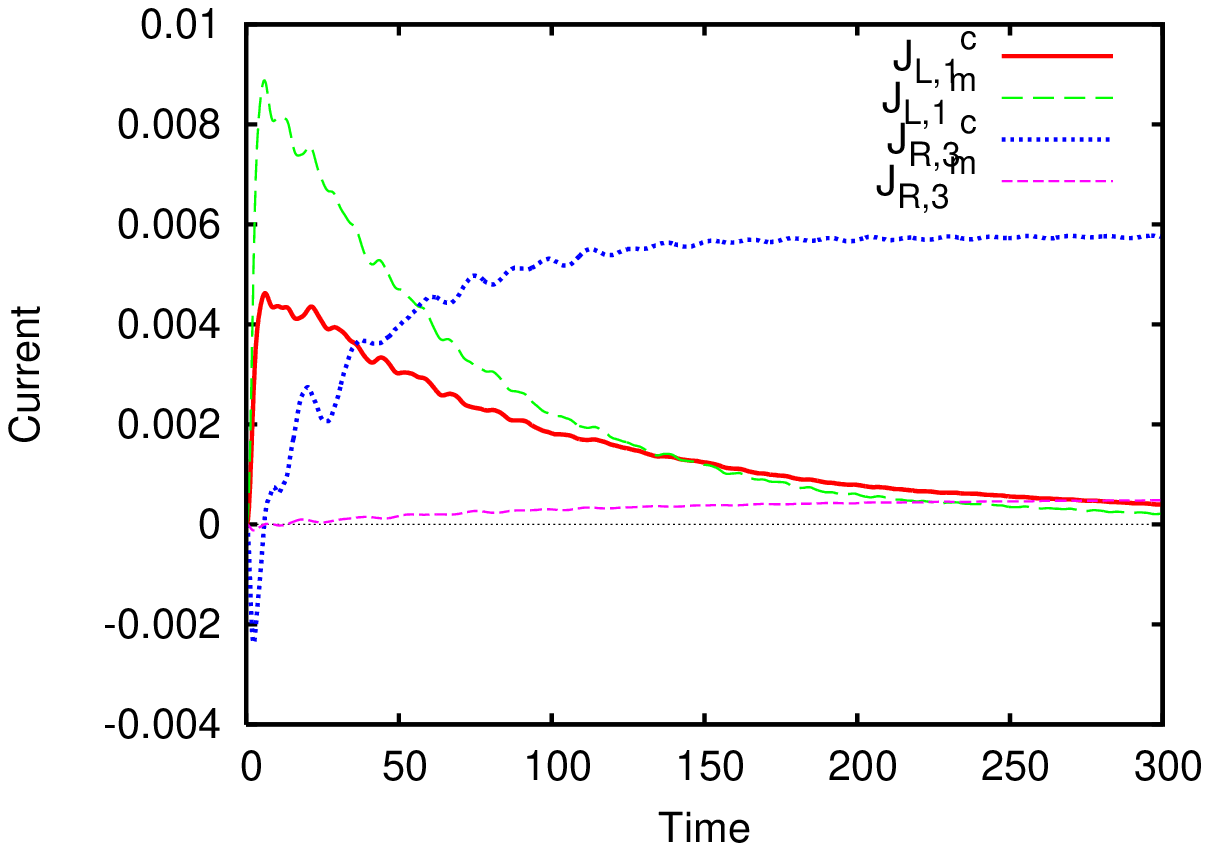}
\end{center}
\caption{(Color online) (a) The coupling coefficients $|T_{qn}|^2$ 
in the case of coupling at the middle sites 
and at opposite corner sites.(b) Comparison of the current carried by the 1st and 3rd levels from
the active region for two configurations of the contacts (see the discussions in the text).
}
\label{figure9}
\end{figure}

\begin{figure}[tbhp!]
\begin{center}
\includegraphics[width=0.4\textwidth]{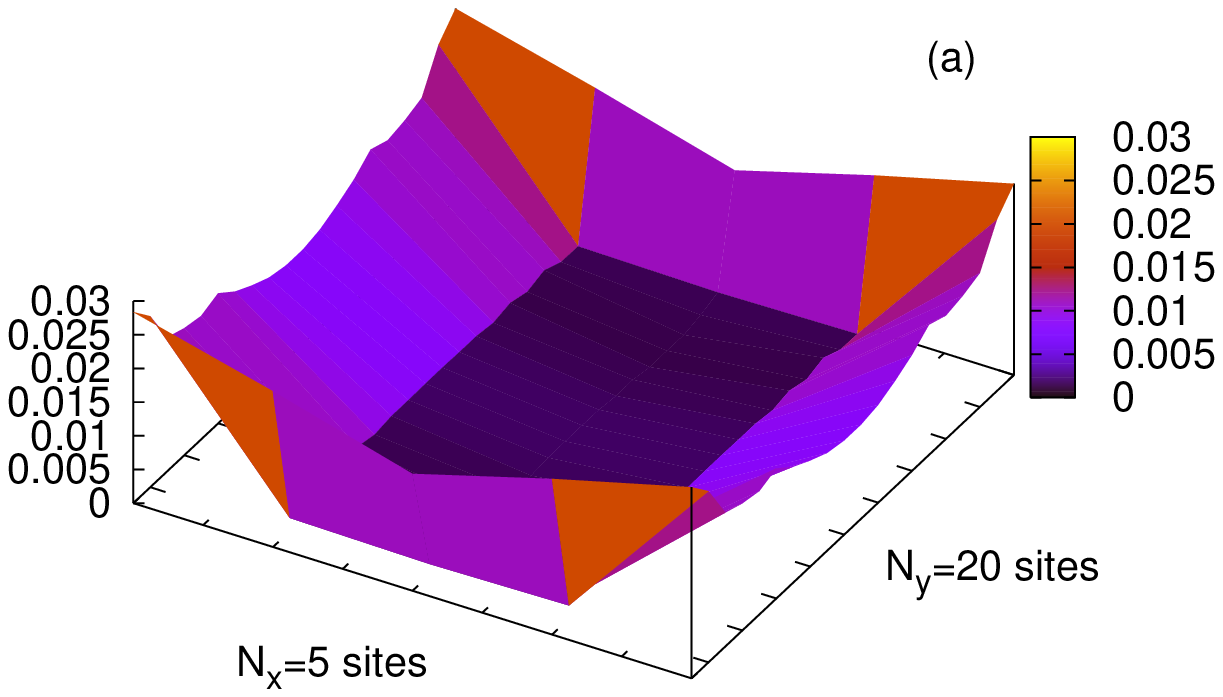}
\vskip -1cm
\includegraphics[width=0.4\textwidth]{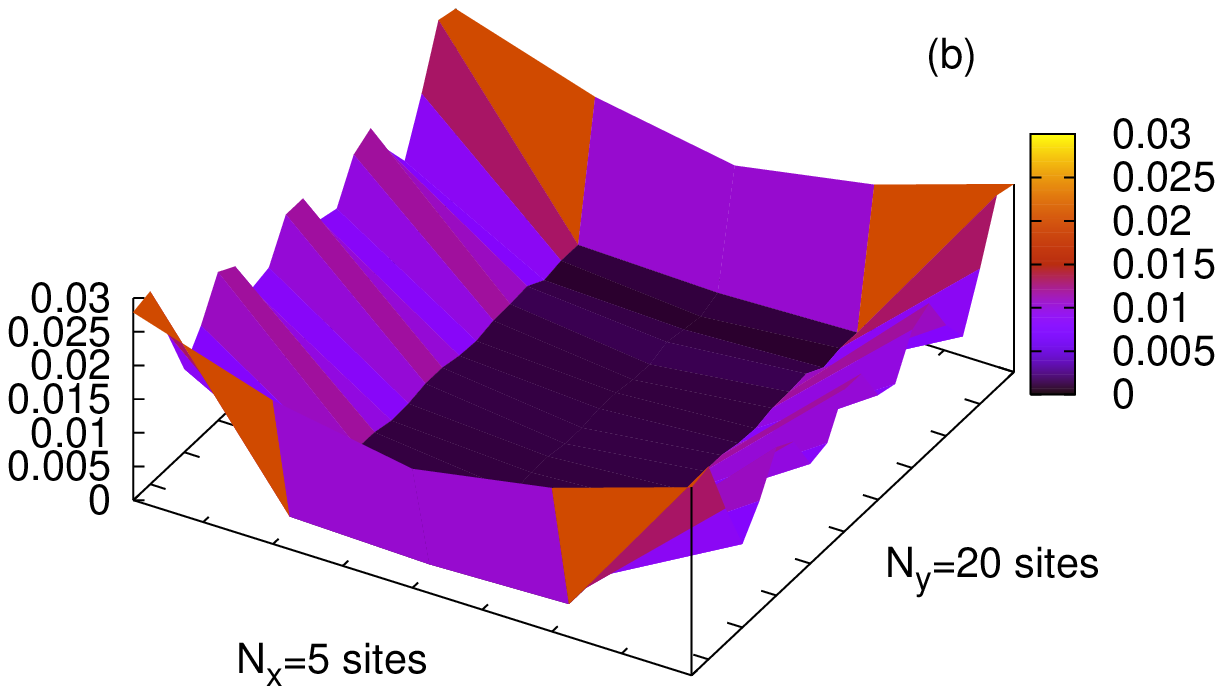}
\end{center}
\begin{center}
\includegraphics[width=0.45\textwidth]{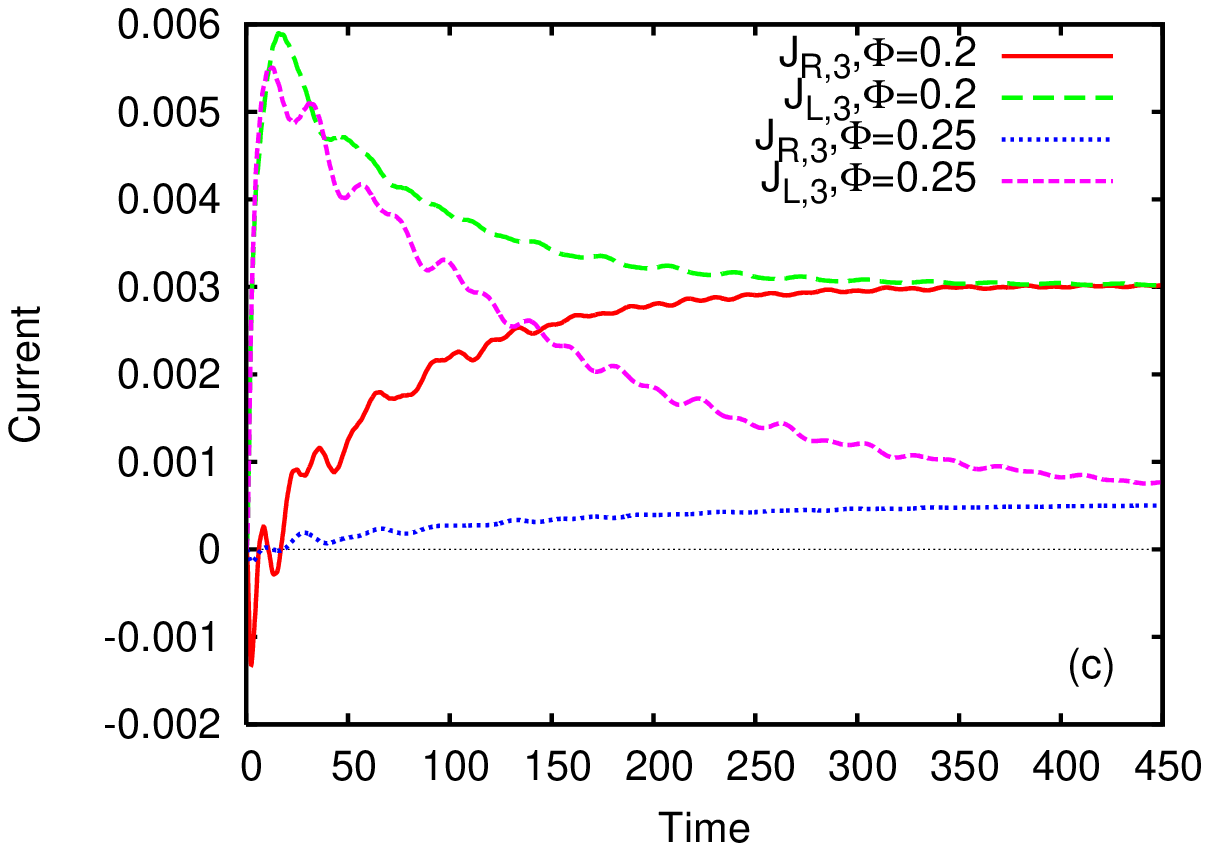}
\includegraphics[width=0.45\textwidth]{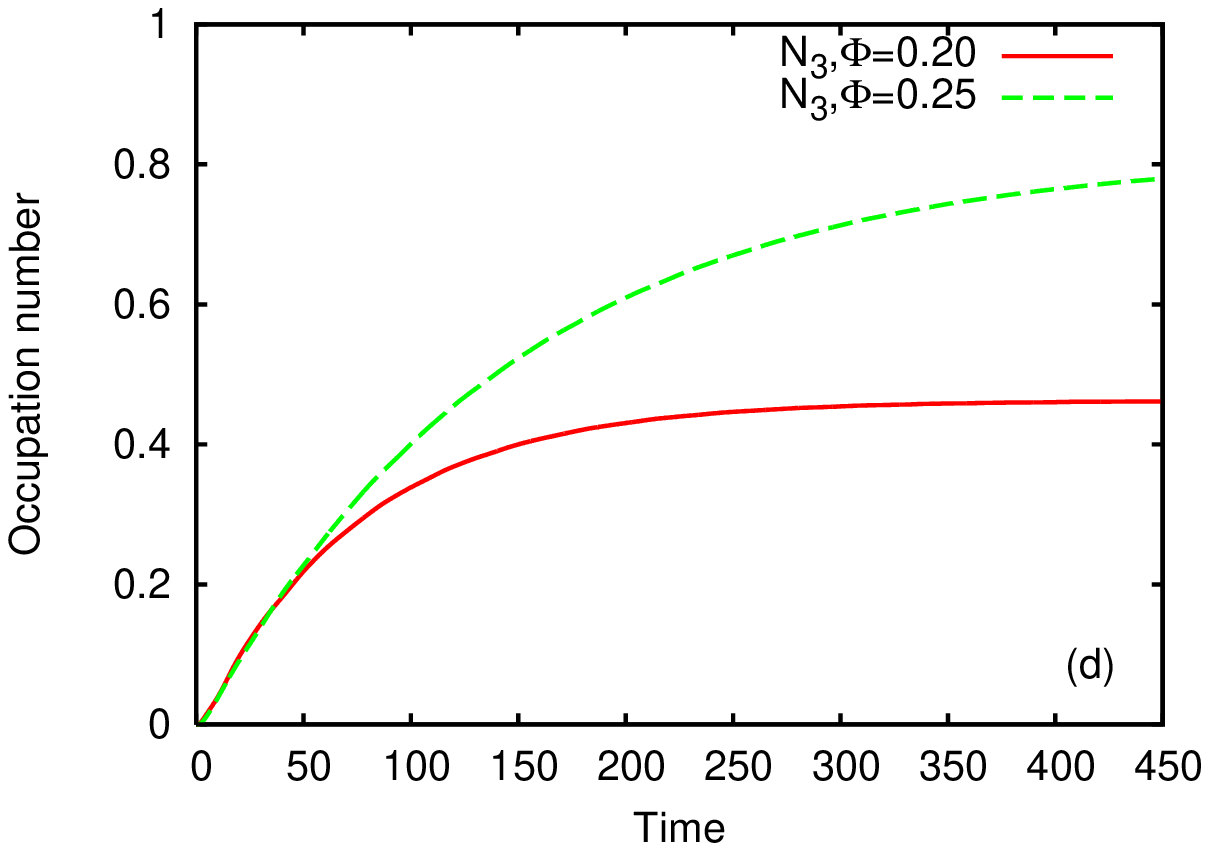}
\end{center}
\caption{(Color online) (a) The on-site probability $|\phi_{3}(i)|^2$
for a $5\times 20$ sites sample at a magnetic flux $\Phi=0.2$
 (b) The same for $\Phi=0.25$. Note the oscillatory behavior along the sample.
(c) The currents associated to the states shown in (a) and (b). (d) The occupation number $N_3$ for the
two values of the magnetic field.
Other parameters: $\mu_L=-1.5$ and $\mu_R=-2.25$, $V_L=V_R=1$, $kT=10^{-4}$.}
\label{figure10}
\end{figure}

Another issue we consider in this work is related to the propagation of electrons across the sample.
More precisely, we want to investigate the behavior of transients carried by edge states having similar amplitudes
in the {\it contact} region but different structures {\it along} the sample. Such a comparison can be done
by changing the magnetic flux while keeping the bias window fixed. For example, a $5\times 20$ sites plaquette
with two leads attached to opposite corners has four states located within the bias window at $\Phi=0.2$
and $\Phi=0.25$ if the chemical potentials of the leads are $\mu_L=-1.5$ and $\mu_R=-2.15$.
The energy of the states within the bias window changes with the magnetic field but the tunneling amplitudes
$|T_{qn}|^2$ remain roughly the same (not shown). The main difference occurs in the on-site localization
probability $|\phi_n(i)|^2$, $n=1,..,4$. We show in Fig.\ 10(a) and (b) the localization
of the 3rd state from the bias window for the two values of the magnetic flux.
At the corners, where the leads are now attached, the states are similar, but
the state corresponding to $\Phi=0.25$ has many oscillations along the sample edge. The question is then:
is there any difference in the transient currents carried by the two states? Fig.\ 10(c)
shows that the oscillating state in Fig.\ 10(b) is not favorable for transport, as $J_{R,3}$ is very small, even
if there is a net current entering the left lead. In contrast, at $\Phi=0.2$ the 3rd level carries
a substantial current in the right lead. These results suggest that the state given in Fig.\ 10(b) should be
almost filled in the steady-state regime because the electrons are trapped inside the sample after entering from
the left lead. This fact is confirmed by inspecting the occupation number $N_3$ at the two values
of the magnetic flux (see Fig.\ 10(d)). The steady-state value of $N_3\sim 0.5$ for $\Phi=0.2$
 meaning that on average this level is only half-occupied and allows the propagation of electrons in the right lead,
whereas at $\Phi=0.25$ the level is more filled and therefore the associated current $J_{R,3}$ is very small.
Note also that for $\Phi=0.25$ the steady-state is achieved very slowly.

\begin{figure}[tbhp!]
\begin{center}
\includegraphics[width=0.4\textwidth]{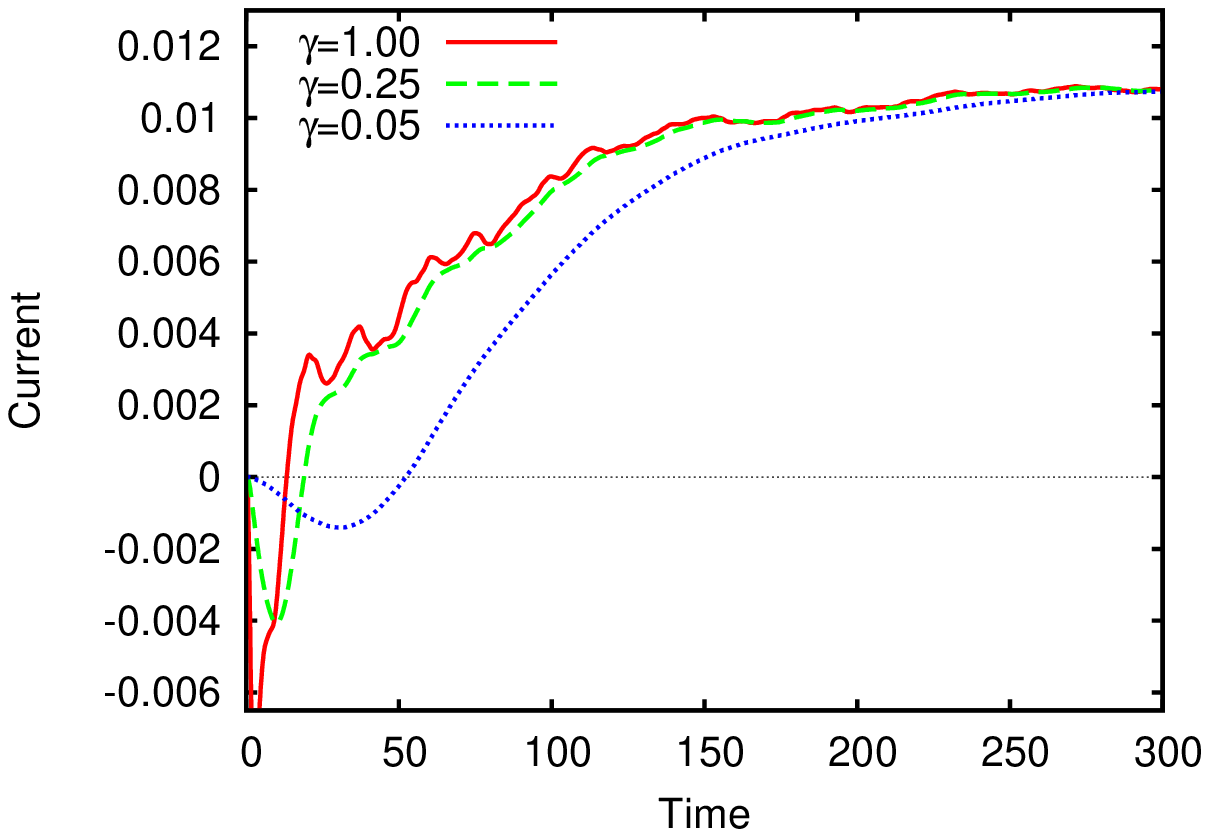}
\vskip -3cm
\hskip 2.75cm
\includegraphics[width=0.15\textwidth]{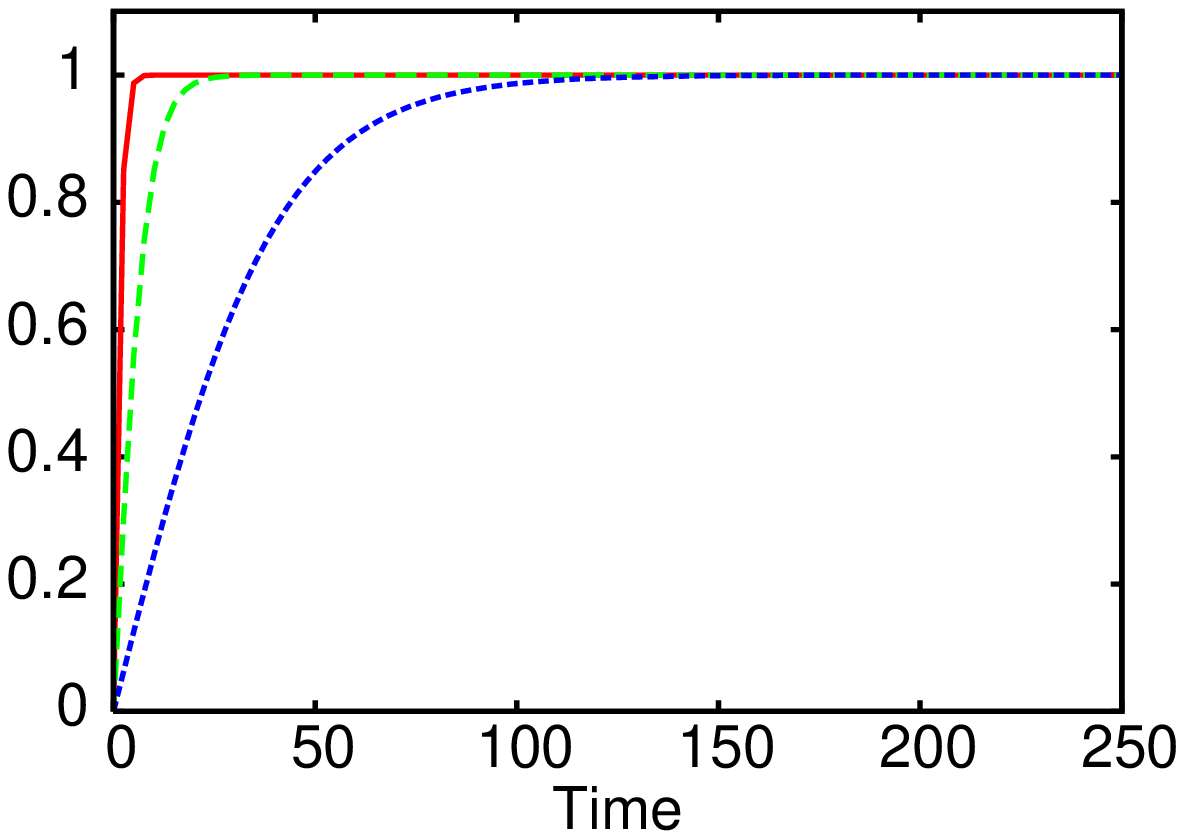}
\vskip 0.75cm
\end{center}
\caption{(Color online) The total current $J_R$ for a $5\times 10$ sites sample and for
fast, moderate and slow coupling to the leads. The inset shows the switching functions
corresponding to the three currents.
Other parameters: $\rho(0)=|{\bf 1}\rangle\langle {\bf 1}|$, $\Phi=0.2$, $\mu_L=-1.5$
and $\mu_R=-2.25$, $V_L=V_R=1$, $kT=10^{-4}$.}
\label{figure11}
\end{figure}

We have shown up to now that the transients depend on the position of the contacts, on the
initial configuration of occupation numbers in the sample and also on the localization properties
of the carrying state. To complete our analysis we shall finally investigate the dependence of
the transport properties on the switching functions $\chi^l$. More precisely, we shall keep
the form of $\chi^l$ as introduced in the beginning of this section and compute the
currents for different values of the parameter $\gamma$. Fig.\ 11 gives the total current
in the right lead for the $5\times 10$ sample with $\rho(0)=|{\bf 1}\rangle\langle {\bf 1}|$ and
with the same parameters as in Fig.\ 4. It is clear that as $\gamma$ decreases the filling of the
systems from the right lead slows down and $J_R$ decreases. For $\gamma =0.05$, which corresponds
to a very slow coupling to the leads, we notice a delay even in the filling process (i.e. $J_R$
is vanishingly small for $t<20$).
 Also, the small oscillations of the transient that appear at
$\gamma=1$ are reduced at $\gamma=0.25$ and disappear as the coupling is established even slower.
Another interesting fact we could learn from Fig.\ 11 is that the steady-state current does
not depend on the switching function. This feature is intuitively expected and was even
rigorously proved by Cornean {\it et al.} \cite{Horia}

\begin{figure}[tbhp!]
\begin{center}
\includegraphics[width=0.4\textwidth]{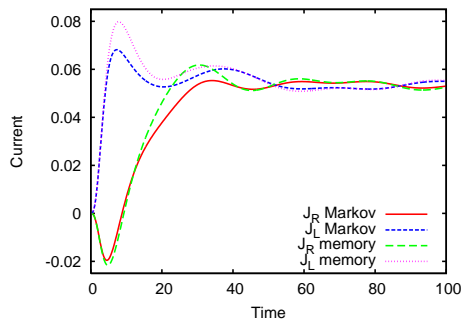}
\end{center}
\caption{(Color online) The total currents $J_{R,L}$ computed from both Markov
approximation and the non-Markovian versions of GME for a $5\times 25$ sites sample.
Other parameters: $\rho(0)=|{\bf 1}\rangle\langle {\bf 1}|$, $\Phi=0.2$, $\mu_L=1.7$
and $\mu_R=1.35$, $V_L=V_R=1$, $kT=10^{-4}$, $\gamma=0.75$.}
\label{figure12}
\end{figure}

A natural step forward in our analysis would be to look for memory effects, that is, to compare
the solution of the GME with the Markovian case. Non-Markovian effects are expected to appear
for strong lead-sample coupling or if one has entanglement in the initial configuration.
Usually, the memory effects can be neglected if the dynamics of the reservoirs is much faster
than the dynamics of the sample. In our case we cannot check this fact because the relevant
time scales (i.e.\ the correlation time in the leads and the dynamics of the sample) are not at hand for the complex systems we study. They presumably depend on the bias, on various energy gaps
and even on the switching functions. This is actually one reason to start from the very beginning
with a non-Markovian GME. Although a closer investigation of memory effects is left for
future work we shall give here a brief account on the Markovian version of Eq.\ (\ref{GME}) and a first
comparison between the two approaches.
Following Timm \cite{Timm} we replace $U_0(t,s)\rho(s)U_0(t,s)^{\dagger}$ by $\rho(t)$ in Eq.\ (\ref{GME})
and compute the $\Pi$
operator with this ansatz. Note that we do not extend the upper integration limit to $\infty $
as is done in most papers but rather compute numerically the integral.

We have performed numerical simulations for a $5\times 25$ sites system, taking just two single particle states within the bias window. The total Markovian and non-Markovian currents $J_{L,R}$
in each lead  are shown in Fig.\ 12. We observe that the memory effects appear at $t\sim 7.5$ and that the currents coincide in the long-time limit $t>45$ (the switching function reaches its maximum value quite fast, around $t\sim 6$). Note also that the memory effects develop
first in $J_L$ and only later in $J_R$. This preliminary calculation shows that indeed the transient currents may not be suitably captured within the Markov approximation, while the steady-state
regime is satisfactorily described. Further investigation is needed in order to understand
memory effects in more complicated processes, e.g.\ pumping.

 Finally we consider as well a two-level system which is extensively studied both within GME and rate equation 
approach. The central region is composed of only two sites to
which the two leads are attached. The two levels are symmetric w.r.t. zero, i.e. $E_1=-1, E_2=1$.
We have taken a bias window that covers both levels $\mu_L=1.25, \mu_R=-1.25$. The transients
show damped oscillations similar to the ones presented by Gurvits {\it et al.} \cite{GG}
 In the steady state regime the two states carry equal currents. The lowest level oscillates more and its associated current
is slightly negative at short times due to the back-tunneling processes from the right lead to the
sample. This feature dissapears when the chemical potential of the right lead is lowered to $\mu_R=-2$,
because the back-tunneling processes are suppresed (not shown). 

\begin{figure}[tbhp!]
\begin{center}
\includegraphics[width=0.4\textwidth]{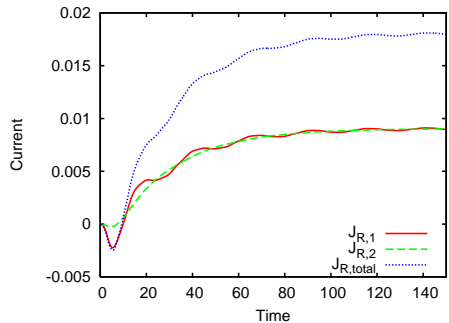}
\end{center}
\caption{(Color online) The currents in the right for a two-level system. 
Other parameters: $\rho(0)=|{\bf 1}\rangle\langle {\bf 1}|$, $\Phi=0.0$, $\mu_L=1.25$
and $\mu_R=-1.25$, $V_L=V_R=0.25$, $kT=10^{-4}$, $\gamma=0.5$.}
\label{figure13}
\end{figure}

\section{Conclusions and discussion}

One of the main features of mesoscopic systems is that their geometry leaves some fingerprints on the
transport properties (e.g.\ the invasive role of the current probes).
In this paper we have described theoretically the time-dependent transport through such structures within
the RDO formalism borrowed from quantum optics. When extended to open quantum systems
this method is a powerful tool for studying electron dynamics through and within a sample coupled to biased
leads and characterized by a well defined initial state. We complement previous approaches
\cite{Gurvitz,Harbola,Pedersen,Vaz} in the following way: i) the GME is solved
without the Markov approximation for arbitrary time-dependent coupling to the leads; ii) the usual
assumption that the spectrum of the sample is entirely contained into the bias window is not needed;
iii) the transfer Hamiltonian describing the coupling between the leads and the sample
takes into account the localization of the sample states
depending on the geometry of the sample and on the region where the leads are plugged.

The GME is solved using the Crank-Nicolson algorithm.
In real experiments the sample is characterized either by a ground state or by a low-energy excited
state and for small couplings to the leads one expects that most of the
levels below (above) the bias window remain occupied (empty) and will not contribute to the current.
The relevant many-body states are actually few and
they are given by all combinations of occupation numbers for a bunch of single particle states from the vicinity
of the bias window. Motivated by this fact one can actually restrict the calculation of the matrix elements
for the RDO. A refined version of the argument should hold in the presence of the
Coulomb interaction as well, using perhaps a Hartree-Fock initial ground state.

The numerical simulations are obtained for a lattice Hamiltonian. As a main application of the method we have computed the transients
associated to each level of a two-dimensional lattice in the presence of a strong perpendicular magnetic
field. In this case the currents are carried only by edge states and depend both on the contact point and
on the topology of the state. Different initial states of the isolated system lead to different transients
but to the same steady-state current which is not achieved at the same time. We have presented a comprehensive
analysis of the electron dynamics in the transient regime and also studied the relevant matrix elements
of the RDO (both populations and coherences are discussed). Also we have shown that the 
Markov approximation is appropriate for steady-state calculation but misses some memory effects in the transient regime. We want to emphasize that our method not only goes beyond the Markov
and wide-band approximations, being thus from the very beginning more
accurate, but it is even more efficient for numerical calculations:
the time integration can be done recursively, which is not possible in
the Markov approximation.  Consequently the calculations in the Markov
approximation took much longer computer time than the solution of the GME.

Some of the transient properties, like the delay in the appearance of a current in the drain lead
should motivate further experiments in the field. Further expected applications of this method include
pulse propagation, pumping and time-dependent interference effects.

The electron-electron interaction inside the central region was not
included in the present calculations and therefore subtle features
like charging effects or charging effects on the displacement currents
could not be discussed here. In the case of many-level systems with
a specific geometry the treatment of Coulomb interaction within a
time-dependent framework is a highly non-trivial issue.  In particular,
the exact diagonalization method (mostly used for two-level systems in the
cotunneling regime) becomes numerically costly due to the large number of
many-body states. Therefore the calculation of the {\it time-dependent}
number of particles in the presence of electron-electron interactions
requires a suitable approximation scheme. We believe however that our
approach provides a faithful qualitative description of the main dynamical
processes and gives a first hint about the crucial role played by the
spectral properties of many-level systems in the transient regime.

\appendix
\section{Derivation of GME}

We present here for completeness the derivation of Eq.\ (\ref{GME}) following essentially the
Nakajima-Zwanzig method. To this end we write the equation of motion for $W(t)$ in terms of the
Liouvillian ${\cal L}$:
\begin{eqnarray}\label{Liouv}
i\hbar\frac{dW(t)}{dt}&=&{\cal L}(t)W(t),\quad {\cal L}(t)={\cal L}_0+{\cal L}_T(t)\\
{\cal L}_0\cdot &=&[H_0,\cdot],\quad {\cal L}_T(t)=[H_T(t),\cdot].
\end{eqnarray}
Next we define two projections:
\begin{equation}
P\cdot=\rho_L\rho_R{\rm Tr}_L{\rm Tr}_R\{\cdot \}\quad Q=1-P.
\end{equation}
It is straightforward to check the following properties:
\begin{eqnarray}
P{\cal L}_S&=&{\cal L}_SP,\quad P{\cal L}_T(t)P=0.
\end{eqnarray}
The Liouville equation (\ref{Liouv}) splits then into two equations:
\begin{eqnarray}\label{PW}
i\hbar P\dot W(t)&=&P{\cal L}(t)PW(t)+P{\cal L}(t)QW(t)\\
i\hbar Q\dot W(t)&=&Q{\cal L}(t)QW(t)+Q{\cal L}(t)PW(t),
\end{eqnarray}
and the second equation can be solved by iterations ($T$ is the time-ordering operator):
\begin{equation}\label{QW}
QW(t)=\frac{1}{i\hbar}\int_{t_0}^tdsT\exp \left\lbrace
-\frac{i}{\hbar}\int_s^tds'Q{\cal L}(s') \right\rbrace
Q{\cal L}(s)PW(s).
\end{equation}
Inserting Eq.\ (\ref{QW}) in Eq.\ (\ref{PW}) and using the properties of $P$ we get the following
equation:
\begin{eqnarray}\nonumber
\hskip -2cm i\hbar P{\dot W}(t)=P{\cal L}_SW(t)\\\label{PdotW}
\hskip -0.25cm
+\frac{1}{i\hbar}P{\cal L}_T(t)Q\int_{t_0}^tds
T\exp \left\lbrace
-\frac{i}{\hbar}\int_s^tds'Q{\cal L}(s')Q \right\rbrace Q{\cal L}_T(s)PW(s).
\end{eqnarray}
In order to have an explicit perturbative expansion in powers of the transfer
 Hamiltonian $H_T(t)$ one has to factorize the time-ordered exponential as:
%unperturbed part $\exp Q{\cal L}_0Q$
in the time-ordered exponential in Eq.\ (\ref{PdotW}):
\begin{equation}
T\exp \left\lbrace
-\frac{i}{\hbar}\int_s^tds'Q{\cal L}(s')Q \right\rbrace =\exp \{Q{\cal L}_0Q\}
(1+{\cal R}),
\end{equation}
where the remainder ${\cal R}$ contains higher powers of $H_T$. The Born approximation
of the generalized master equation consists in neglecting ${\cal R}$. Another technical point
is that in expanding the unperturbed part $\exp Q{\cal L}_0Q$ one can replace each $Q$ between two
Liouvillian ${\cal L}_0$ by $P+Q=1$, since $Q{\cal L}_0P{\cal L}_0Q=0$. By taking
the trace over the leads in Eq.\ (\ref{PdotW}) one obtains:
\begin{equation}\nonumber
i\hbar{\dot\rho}={\cal L}_S\rho(t)
+\frac{1}{i\hbar}{\rm Tr}_L{\rm Tr}_R\{{\cal L}_T(t)
\int_{t_0}^tdse^{-i(t-s){\cal L}_0}{\cal L}_T(s)\rho_L\rho_R\rho (s)   \},
\end{equation}

which is nothing but Eq.\ (\ref{GME}).

\section*{Acknowledgments}This work was supported in part by the Icelandic Science and Technology Research
Program for Postgenomic Biomedicine, Nanoscience and Nanotechnology, the Icelandic Research Fund, and the Research
Fund of the University of Iceland. V.M acknowledges the hospitality of
 the Science Institute where this work was initiated and the financial support from PNCDI2 programme.

\end{document}